\documentclass[a4paper]{llncs}
\usepackage{fullpage}

\usepackage{color}
\usepackage[]{algorithmic}
\usepackage{algorithm}
\usepackage{subfig}
\usepackage{pgf}
\usepackage{tikz}
\usetikzlibrary{arrows,automata,shapes,shadows,calc}
\usetikzlibrary{positioning}
\usepackage{amsmath,amssymb}
\usepackage{xspace}
\usepackage{tabularx}
\usepackage{colortbl}
\usepackage{nicefrac}
\usepackage{stackengine}
\usepackage{setspace}
\usepackage{picins}
\usepackage{graphicx}
\usepackage{bm}

\usepackage{listings}
\lstdefinestyle{my_style}{
    numberstyle=\tiny,
    basicstyle=\footnotesize,
    breakatwhitespace=false,
    breaklines=false,
    keepspaces=true,
    numbers=left,
    numbersep=5pt,
    showspaces=false,
    showstringspaces=false,
    showtabs=false,
    tabsize=4,
    frame=single,
    xleftmargin=0.415cm,
    xrightmargin=0.13cm,
    numberbychapter=false
}

\usepackage{paralist}

\usepackage{todonotes}
\newcommand{\ws}{\Sigma^{\#}}

\newcommand{\D}{\ensuremath{{\mathcal D}}}

\newcommand{\satMinD}{\mathsf{sat_{\varphi_{sdm}}}}

\newcommand{\satMin}{\mathsf{sat_{\varphi_m}}}

\newcommand{\satNonMin}{\mathsf{sat_{\overline\varphi_m}}}
\newcommand{\inpExh}{\ensuremath{\mathrm{in\mbox{-}ex}}}

\newcommand{\calP}{{\mathcal P}}
\newcommand{\exec}{\mathit{exec}}

\newcommand{\remove}[1]{}

\newcommand{\ie}{{i.e.,}}

\newcommand{\pref}{\preccurlyeq}

\newcommand{\range}{\ensuremath{\mathrm{range}}}

\newcommand{\bbb}{\mathbb{B}}
\newcommand{\bbn}{\mathbb{N}}

\newcommand{\true}{\ensuremath{\mathsf{true}}}
\newcommand{\false}{\ensuremath{\mathsf{false}}}

\newcommand{\calF}{{\mathcal F}}

\newcommand{\calDF}{{\mathcal {DF}}}
\newcommand{\calDP}{{\mathcal {DP}}}

\newcommand{\calL}{{\mathcal L}}

\newcommand{\eqdef}{\stackrel{{\scriptstyle\mathrm{def}}}{=}}

\makeatletter
\ifx\proof\undefined
\newenvironment{proof}[1][\protect\proofname]{\par
\normalfont\topsep6\p@\@plus6\p@\relax
\trivlist
\itemindent\parindent
\item[\hskip\labelsep\scshape #1]\ignorespaces
}{%
\endtrivlist\@endpefalse
}
\providecommand{\proofname}{Proof}
\fi

\makeatother

\title{Monitoring Data Minimisation \thanks{This work has been partially supported by the Swedish Research Council (grant Nr. 2015-04154, PolUser: Rich User-Controlled Privacy Policies).}}
\author{Srinivas Pinisetty \inst{1}, Thibaud Antignac \inst{4}, David Sands \inst{2}, Gerardo Schneider\inst{3}}
\institute{University of Gothenburg, Sweden \quad \email{srinivas.pinisetty@gu.se} \and
Chalmers University of Technology, Sweden \quad \email{dave@chalmers.se}\and
University of Gothenburg, Sweden, \quad \email{gerardo@cse.gu.se } \and
CEA List, Software Security Laboratory, F-91191 Gif-sur-Yvette Cedex, France \quad\email{thibaud.antignac@cea.fr}
}

\begin{document}
\maketitle
\pagestyle{plain}
\begin{abstract}
Data minimisation is a privacy enhancing principle, stating that personal data collected should be no more than necessary for the specific purpose consented by the user.
Checking that a program satisfies the data minimisation principle is not easy, even for the simple case when considering deterministic programs-as-functions.
In this paper we prove (im)possibility results concerning runtime monitoring of (non-)minimality for deterministic programs both when the program has one input source (monolithic) and for the more general case when inputs come from independent sources (distributed case).
We propose monitoring mechanisms where a monitor observes the inputs and the outputs of a program, to detect violation of data minimisation policies. 
We show that monitorability of (non) minimality is decidable only for specific cases, and
detection of satisfaction of different notions of minimality in undecidable in general. 
That said, we show that under certain conditions monitorability is decidable and we provide an algorithm and a bound to check such properties in a pre-deployment controlled environment, also being able to compute a minimiser for the given program.
Finally, we provide a proof-of-concept implementation for both offline and online monitoring and apply that to some case studies.
\end{abstract}

\section{Introduction}
\label{sec:intro}
According to the Article 5 of the \emph{General Data Protection Regulation} proposal (GDPR ---EU--2016/679, adopted on 27 April 2016 and entering into application on 25 May 2018), ``Personal data must be adequate, relevant, and limited to the minimum necessary in relation to the purposes for which they are processed''~\cite{gdpr2016}.
While determining what is ``adequate'' and ``relevant'' might seem difficult given the inherent imprecision of the terms, identifying what is ``minimum necessary in relation to the purpose'' seems to be easier.  
We could understand this principle in different ways, and we discuss below a couple of possible interpretations. We do so by having in mind that our objective is to find a way to enforce, or at least detect, when the {\em data minimisation} principle is (not) satisfied from a technical point of view, and in particular by using a language-based approach.

One way to understand minimisation is on how the data is {\em used}, that is we could consider ways to identify how the input data is used in the program, for which {\em purposes}. This would imply that we look inside the program and track the usage of the data by performing static analysis techniques like tainting, def-use, information flow, etc. For that we need of course to have a precise definition of what purpose means and a way to check that the intended purpose matches with the real purpose under which data will be processed.

Another way to see minimisation is by considering when and how the data is {\em collected} and only allow the collection of data that is actually needed to compute what is required to achieve the given purpose. In this case we could consider that the ``purpose'' is given by the specification of the program. 

In this paper we take the second view, following~\cite{ASS17dm}.
This kind of data minimisation calls for semantic foundations to determine whether or not a program could run equally well with less personal data input.
Indeed, syntax-driven techniques do not give any information about the semantic ``necessity'' as meant by the proposal.
It is to be noted that this principle exists in other regulations and is sometimes referred to as ``collection limitation'' when it focuses on the particular step of collecting data.
This is for instance the case of the \emph{Fair Information Practice Principles} (FIPPs)~\cite{fipp1973} in USA, and the \emph{Guidelines on the Protection of Privacy and Transborder Flows of Personal Data}~\cite{oecd2013} proposed by the Organisation for Economic Co-operation and Development (OECD).

Determining the quantity of information actually needed for a given purpose requires an analysis of the program.
Following the approach detailed in~\cite{smith2009}, it is possible to quantify the amount of information input to a program, the amount of information semantically used to compute the output, and the amount of input information not semantically used.
If we consider data minimisation from the regulatory point of view, the input data not semantically used in the program should not be collected (and thus not processed). 
This is because, unlike the case in~\cite{smith2009} where the attacker only has access to the outputs, the attacker is here the data processor\footnote{``Data controllers'' and ``data processors'' are legal roles used to define obligations and liabilities of the parties. We indistinctly use the term ``data processor'' in this paper as we are interested in designating the party that technically processes the data.} itself, having then the possibility to also exploit the inputs.
As a consequence, the attacker knows all the information available after the input is collected (before the program execution).

Given that \emph{input data = necessary data + extra data}, and since the program should execute equally well without any extra data, we have that \emph{input data $\geq$ necessary data}.
The goal of the \emph{data minimisation process} is thus to minimise the input data so only what is necessary is given to the program.
Whenever the input data exactly matches what is necessary we may say that the minimisation is \emph{perfect}. Perfect minimisation is, however, difficult to achieve in general among other things because it is not trivial to exactly determine what is the input needed to compute each possible output \cite{ASS17dm}. 
That said, it could be possible to achieve \emph{some degree} of data minimisation, which though not optimal could still be considered useful (we could at least state that the program under consideration does use more input data than needed), or to be able to detect whether the data minimisation principle is violated during execution of the program.

When dealing with data minimisation we could ask ourselves the following two questions. First, ``does this program perfectly respect the data minimisation principle?''. If the answer is Yes, then we are happy and we could certify that the program is in conformance with the regulation. If the answer is No then we should ask ourselves whether the program could be somehow transformed so it satisfies the minimisation principle. Or, instead we could ask ourselves ``is it possible to get a data minimiser such that it generates only the necessary inputs for the given program?'' By trying to answer the latter, instead of trying to transform the original program, a procedure could be given to achieve data minimisation. This is exactly the solution proposed in \cite{ASS17dm} based on the generation of another program (called the \emph{data minimiser}) that filters the input given to the original program so that it is run on a smaller set of input data (and without changing its behaviour).

Let us consider a simple program to exemplify this notion of data minimisation and sketch our solution (see Figure~\ref{lst:benefits_simple_program}).
The purpose of the program is to compute the \emph{benefit level} of employees depending on their salary (assumed to be between \$ $0$ and \$ $100000$).
For the sake of simplicity, in what follows we do not assume any particular distribution over their domain for the inputs, driving the analysis on worst-case assumptions.
A quick analysis of this small program clearly shows that the range of the output is $\left\{ \false, \true \right\}$, and consequently the data processor does not need to precisely know the real salaries of the employees to determine the benefit level. 
In principle each employee should be able to give any number between $0$ and $9999$ as input if they are eligible to the benefits, and any number between $10000$ and $100000$ otherwise, without disclosing their real salaries.

\begin{figure}[t!]
    \centering
\begin{lstlisting}[style=my_style]
input(salary)
benefits := (salary < 10000)
output(benefits)
\end{lstlisting}
    \caption{Running example $P_\textit{bl}$ to compute a benefits level.}
    \label{lst:benefits_simple_program}
    \vspace*{-0.5cm}
\end{figure}

Antignac et.~al.~\cite{ASS17dm} defined the concept of data minimiser as a pre-processor that filters the input of the given program in such a way that the functionality of the program does not change but it only receives data that is necessary and sufficient for the intended computation. From there they derived the concept of data minimisation and they showed how to obtain data minimisers for both the {\em monolithic} case (only one source of input) and the {\em distributed} case (more than one, independent, source of inputs). The latter is clearly a semi-decision procedure given the underlying undecidability result of the problem (it reduces to computing the kernel of a function which semantically denotes the program).
The approach to obtain minimisers is based on a combination of using a symbolic execution engine and a SAT solver. The proof-of-concept implementation provided in that paper only works (automatically) for simple programs (without loop, recursion, nor call to libraries). By providing loop invariants, additional specification for libraries, etc., it is possible to get a semi-automatic way to get minimisers though the manual effort hampers a large scale use of this approach.

Being one of the first papers on data minimisation, we provide a formal definition of minimisation and the study of some of its properties. The approach is however quite limited, mostly in what concerns the generation of data minimisers and even checking whether a given program is minimal or not.

In this paper we take those results further by consider a more practical approach. Knowing that it is in general impossible to compute data minimisers for arbitrary programs by static analysis, we consider here a {\em runtime} approach. 
Our starting point is the definition of data minimisation as a modification of Cohen's notion of {\em strong dependency}~\cite{cohen1977}. As in~\cite{ASS17dm} we consider both the monolithic and the distributed case. We define the notion of runtime monitors for both cases extracting them from the definition of monolithic and (weak) distributed (non-)minimality. 
We consider two different scenarios: 
(i) After the program has been deployed, we perform {\em online monitoring} without any knowledge about how the environment will produce the inputs;
(ii) Before deployment, where we could perform both {\em offline monitoring} (where traces are produced beforehand and fed into the monitor), or {\em controlled online monitoring} where we produce the inputs in a systematic way in order to capture as much of the input domain as possible (all in some cases) in a way reminiscent to the approaches developed for test cases generation.

As for most non-interference properties, checking whether a given program satisfies the data minimality principle implies checking a {\em hyper-property}~\cite{hyp1,Finkbeiner2015}, that is a property over set of traces and not over a single trace. For monitoring hyperproperties we need to consider multiple executions of the program and this has been shown to be computationally hard~\cite{Bonakdarpour2016}.  
In general it is not possible to reduce hyper-property checking into checking over a single trace, but for (non-)minimality this is the case. As a consequence, we transform the problem of checking a hyper-property over a given program into checking a property over repeated executions of the program (program-in-a-loop) so the monitoring problem can be reduced to the analysis of a single trace.

We briefly sketch here our approach over the example shown in Figure~\ref{lst:benefits_simple_program}, focusing only on how to check non-minimality using online monitoring, and how we use a slight variation of the same procedure before deployment to obtain a minimiser\footnote{The example is only for the monolithic case and is simple on purpose. In the rest of the paper we give a formalisation of all the concepts and present results for the more general case also (two versions of distributed minimality).}.
Our monitor is a simple program taking input/output pairs and checking whether for different inputs the program gives the same output. Informally, if the monitor finds two pairs $(i_1,o_1)$ and $(i_2,o_2)$ such that $i_1 \neq i_2$ but $o_1 = o_2$ then this would be a violation of (monolithic) minimality (and such pairs would provide a witness for non-minimality)\footnote{For simplicity we consider here multiple executions of the program, that is, a set of traces of length one.}. For this example, if the monitor observes the following two executions: $(5000,\mathit{\true})$, $(11000,\mathit{\false})$, it cannot conclude anything given that no same output for different inputs has been produced. However, after a third execution with any input value different from 5000 or 11000 is done (e.g., $(6500,\mathit{\true})$) the monitor would raise a flag indicating a violation of minimality occurred.
Two important comments are needed here concerning a solution based on runtime monitoring: 
\begin{inparaenum}[i)]
\item It might happen that the monitor never finds a witness for non-minimality even if the program is non-minimal. This could happen if the executions loop over the same inputs while never violating the property (at runtime a monitor can only act on the real executions of the program), or if the domain and co-domain of the program are too big, eventually needing an unbounded number of executions in order to be able to exactly produce the violating trace.
\item It could happen that the program is non-minimal but there is a client-side minimiser filtering the inputs (for instance by choosing always a representative for each possible output). If this is the case then the monitor would never detect that the program is non-minimal. 
\end{inparaenum}
The latter case shows that what our monitoring approach is in fact doing is to check non-minimality for a {\em composed system} formed by the program (server side) and a potential minimiser (client side).

If we do have additional information about the input domain we can do better. In particular, if the input domain is finite, we can use runtime monitoring in a controlled environment in order to give a definitive Yes/No answer to the (non-)minimality problem. 
We combine the monitor as before but now we generate all possible inputs and check (exhaustively) all input/output pairs. Even better, we have a procedure that computes a minimiser for the given program. For our example, the minimiser is simply a program that generates two different constant numbers depending on whether the real input is less than $10000$ (e.g., $5000$) or bigger than $10000$ (e.g., $15000$). In this case, the program is still non-minimal, but we guarantee that it only receives two different values not disclosing the real input (the composed system becomes minimal).
Briefly, we can compute a minimiser for an arbitrary deterministic program under some reasonable assumptions (in practice most programs operate on bounded domains).

We have here summed-up some of the contributions of our paper and given an example for the simplest case (monolithic). In the rest of the paper we present our results in a more formal manner both for the monolithic and the more complex distributed case. We also give proof-of-concepts implementations for our monitoring approach.

\section{Preliminaries}
\label{sec:prelim}
In this section we revisit basic concepts related to runtime verification (Section~\ref{sec:rv}), and introduce other notations that we will use in the paper (Section~\ref{prelim:mono:dist}).

\begin{quote}
According to \cite{LeuckerS08jlap} ``{\em runtime verification} (RV) is the discipline of computer science that deals with the study, development, and application of those verification techniques that allow checking whether a run of a system under scrutiny satisfies or violates a given correctness property''. 
Checking whether an execution meets a correctness property is typically performed using a {\em monitor}, a program that decides whether the current execution satisfies the given property by outputting either yes/$\true$ or no/$\false$. Formally, when $\|\varphi\|$ denotes the set of valid executions given by a property $\varphi$, RV boils down to checking whether a specific execution is an element of $\|\varphi\|$. Thus, in its mathematical essence, runtime verification answers the word problem, i.e. the problem whether a given word is included in some language.
\end{quote}
In general, the monitor is automatically extracted from the property $\varphi$ (which the monitor should verify).
The system being monitored could be considered as a \emph{black-box}, as for instance done by the RV approaches in~\cite{Havelund2008,Bauer:2011:RVL,Blech2012,DBLP:conf/setss/Leucker16,FalconeFM09}, or as a \emph{white-box} (or \emph{grey-box}) as in~\cite{ColomboPS08}.
Properties which the monitor should verify are usually specified in high-level formalisms with a semantics customised for finite executions such as automata theory~\cite{ColomboPS08,FalconeFM09} or some variant of Linear Temporal Logic (LTL) such as LTL$_3$~\cite{Bauer:2011:RVL}.

\begin{figure}[htb]
	\begin{centering}
		\scalebox{0.85}{
			\includegraphics[scale=1]{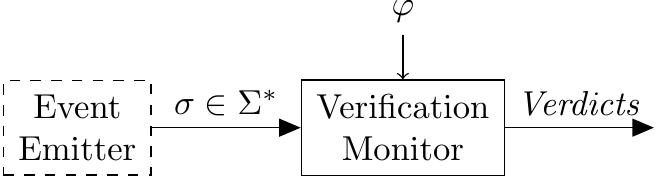}
		}
		\caption{Runtime Verification.}
		\label{fig:rv-mon}
	\end{centering}
\end{figure}

A verification monitor does not influence or change the program execution.
Such a monitor can be used to check the current execution of a system ({\em online}) or a stored execution of a system ({\em offline}).
An execution of a system is considered as a finite sequence of actions emitted by a system being monitored.
As illustrated in Figure~\ref{fig:rv-mon}, a verification monitor for a given property $\varphi$ takes a sequence of events $\sigma$ from a \emph{black-box} system (event emitter) as input and produces a verdict as output that provides information about whether the current execution of the system $\sigma$ satisfies  $\varphi$ or not.
In order to reason about runtime monitoring and verification, we rely on program execution traces. We thus need to introduce some additional basic notions before formally defining a runtime monitor.

A finite word over a finite alphabet $\Sigma$ is a finite sequence $\sigma = a_1\cdot a_2\cdots a_n$ of elements of $\Sigma$.
The set of all finite words over $\Sigma$ is denoted by $\Sigma^*$, and $\Sigma^\#$ denote a subset of $\Sigma^*$ (i.e., $\Sigma^\# \subseteq \Sigma^*$).
The {\em length} of a finite word $\sigma$ is denoted by $|\sigma|$.
The empty word over $\Sigma$ is denoted by $\epsilon_\Sigma$, or $\epsilon$ when clear from the context.
The {\em concatenation} of two words $\sigma$ and $\sigma'$ is denoted as $\sigma\cdot \sigma'$.
A word $\sigma'$ is a {\em prefix} of a word $\sigma$, denoted as $\sigma' \pref \sigma$, whenever there exists a word $\sigma''$ such that $\sigma = \sigma'\cdot \sigma''$; and $\sigma' \prec \sigma$ if additionally $\sigma' \neq \sigma$; conversely $\sigma$ is said to be an \emph{extension} of $\sigma'$.

Given an $n$-tuple of symbols $e=(e_1,\ldots, e_n)$, for $i\in[1,n]$, $\Pi_i(e)$ is the projection of $e$ on its $i$-th element, {\ie} $\Pi_i(e)\eqdef
e_i$.
Given a word $\sigma$ of length $n$, for any $i\in[1,n]$, $\sigma_{i}$ denotes $i^{th}$ element in $\sigma$.
\subsection{Runtime verification monitor }
\label{sec:rv}
In this section, 
we present a definition of a monitor for any given property $\varphi$, and present and discuss some important constraints that it satisfies.

A monitor is a device that reads/observes a finite trace (an execution of the system being monitored) and emits a verdict regarding satisfaction of a given property $\varphi$.
The verdicts provided by the monitor belong to the set $\D= \{ \top, \bot, ? \}$, where verdicts $\true$ ($\top$) and $\false$ ($\bot$) are conclusive verdicts while unknown ($?$) is an inconclusive verdict.
A monitor for any given property $\varphi$ is denoted as $M_{\varphi}$ .
Let us see a definition of a verification monitor for any given property $\varphi \subseteq \Sigma^*$.
%
%
\begin{definition}[RV monitor]
	\label{def:rv:mon}
	Let $\sigma \in \Sigma^*$ denote a current observation of an execution of the system, and
	consider a property $\varphi\subseteq \Sigma^*$.
	A {\em monitor} is a function $M_{\varphi}: \Sigma^* \rightarrow \D$, where $\D= \{\top, \bot, ? \}$ defined as follows:
	\[
	\begin{array}{lll}
	M_{\varphi}(\sigma) & =
	\begin{cases}
	\top &
	\mbox{if }\ \forall \sigma' \in\Sigma^*: \sigma\cdot\sigma' \in \varphi \\
	\bot &  \mbox{if }\ \forall \sigma' \in \Sigma^*: \sigma\cdot\sigma' \not\in \varphi \\
	? & \text{otherwise }
	\end{cases}
	\end{array}
	\]
\end{definition}
%
Property $\varphi$ is a set of finite words over alphabet $\Sigma$ (i.e., $\varphi \subseteq \Sigma^*$).
Verdicts \emph{true} ($\top$) and \emph{false} ($\bot$) are conclusive verdicts, and verdict \emph{unknown} ($?$) is an inconclusive verdict.\footnote{Inconclusive verdict unknown ($?$), can be refined,  where the monitor can provide information/verdict only about the execution seen so far.}
\begin{itemize}
	\item $M_{\varphi}(\sigma)$ returns $\top$ if for any continuation $\sigma'\in \Sigma^*$, $\sigma\cdot\sigma'$ \emph{satisfies} $\varphi$.
	\item $M_{\varphi}(\sigma)$ returns $\bot$ if for any continuation $\sigma'\in \Sigma^*$, $\sigma\cdot\sigma'$ \emph{falsifies} $\varphi$.
	\item  $M_{\varphi}(\sigma)$ returns unknown ($?$) otherwise. 
\end{itemize}
%
\begin{remark}[Monitorability]
	\label{rem:monitorability}
	A property $\varphi\subseteq \Sigma^*$ is {\em monitorable} \cite{Pnueli2006,Bauer:2011:RVL,Falcone2009} if for any observed word $\sigma \in \Sigma^*$, there exists a finite word $\sigma'\in\Sigma^*$ such that the property $\varphi$ can be positively or negatively evaluated for $\sigma\cdot \sigma'$. 
	That is,   $\forall \sigma \in \Sigma^*, \exists \sigma' \in \Sigma^*: M_{\varphi}(\sigma\cdot \sigma') \in \{ \top, \bot\}$.
	All safety (resp. co-safety) properties are monitorable \cite{Bauer:2011:RVL,Falcone2009}.
	For  a safety (resp. co-safety) property,  a monitor can provide a conclusive verdict $\bot$ (resp. $\top$) when it observes a finite word that violates (resp. satisfies) the property.
	It is shown that safety and co-safety properties represent only a subset of monitorable properties \cite{Bauer:2011:RVL,Falcone2009,DIEKERT201429}. 
	Monitorable properties according to safety-progress classification of properties has been discussed in \cite{Falcone2009}, where
	it is shown that Boolean combinations of safety and co-safety properties are monitorable.
	There are some response properties,
	such as ``Every request is acknowledged'', which are non-monitorable since for all finite words, it is never possible to decide satisfaction or violation of the property. This is since  every finite word can be extended to a word that belongs to the property or to a word that does not belong to the property. 
\end{remark}
\begin{proposition}
	For any given property $\varphi \subseteq \Sigma^*$ that is monitorable, monitor  $M_\varphi$ as per Definition \ref{def:rv:mon} satisfies the following constraints:
	
	  \noindent
	  {\bf Impartiality}
	  $\forall \sigma \in \Sigma^* $,
	  \begin{equation}
	  \tag{\bf Imp}\label{eq:snd}
	  \begin{array}{ll}
	  M_{\varphi} (\sigma) =\ ? \text{ iff } \\
	  (\sigma \in\varphi   \wedge \exists \sigma' \in \Sigma^*: \sigma \cdot \sigma' \not\in \varphi) \vee 
	   (\sigma \not\in \varphi   \wedge \exists \sigma' \in \Sigma^*: \sigma \cdot \sigma' \in \varphi)  
	  \end{array}
	  \end{equation}
	  \noindent
	  {\bf Anticipation}
	  $\forall \sigma \in \Sigma^* $,
	  \begin{equation}
	  \tag{\bf Acp}\label{eq:opt}
	  \begin{array}{ll}
	  M_{\varphi} (\sigma) = \top\;  \text{ iff} & (\forall \sigma' \in \Sigma^*: \sigma \cdot \sigma' \in \varphi) \\ 
	  M_{\varphi} (\sigma) = \bot\;  \text{ iff} & (\forall \sigma' \in \Sigma^*: \sigma \cdot \sigma' \not\in \varphi) 
	  \end{array}
	  \end{equation}  
\end{proposition}
%

\emph{Impartiality} expresses that for a finite trace $\sigma \in \Sigma^*$, 
if $\sigma$ is consistent with $\varphi$, but there is some extension of $\sigma$ which is not, or conversely, if $\sigma$ is not consistent with $\varphi$ but some extension is, then the monitor must give verdict $?$ on $\sigma$.

\emph{Anticipation} states that for a finite trace $\sigma \in \Sigma^*$,  the monitor $M_\varphi(\sigma)$ should provide a conclusive verdict $\top$ (resp. $\bot$) iff every continuation of $\sigma$ satisfies (resp. violates)  $\varphi$. 
Thus, anticipation also means that if $M_\varphi(\sigma)$ is $\top$ (resp. $\bot$), then every continuation of $\sigma$ also evaluates to $\top$ (resp. $\bot$).
Formally for any $\sigma\in \Sigma^*$, 
\[ 
\begin{array}{ll}
	(\text{ if }  M_{\varphi} (\sigma) = \top \text{ then } (\forall \sigma' \in \Sigma^*, M_{\varphi} (\sigma \cdot \sigma') = \top) \textbf{ and } \\
	\quad (\text{ if } M_{\varphi} (\sigma) = \bot  \text{ then } (\forall \sigma' \in \Sigma^*, M_{\varphi} (\sigma \cdot \sigma') = \bot)).
\end{array}
\] 

Constraints \ref{eq:snd} and \ref{eq:opt} ensure that the monitor provides a conclusive verdict as soon as possible.
That is, constraints \ref{eq:snd} and \ref{eq:opt} also ensure the following:
			\begin{itemize}
				\item
				\[
				\begin{array}{ll}
				\forall \sigma \in \Sigma^* , 
				(M_{\varphi} (\sigma) = \bot \wedge \forall \sigma' \prec \sigma, M_{\varphi} (\sigma') = ?) \\ 
				\quad \quad \quad \implies
				\forall \sigma' \prec \sigma, \exists \sigma'' \in \Sigma^*: \sigma' \cdot \sigma^{''} \in \varphi.
				\end{array}
				\]
				\item
				\[
				\begin{array}{ll}
				\forall \sigma \in \Sigma^* , 
				(M_{\varphi} (\sigma) = \top \wedge \forall \sigma' \prec \sigma, M_{\varphi} (\sigma') = ?) \\ 
				\quad \quad \quad \implies
				\forall \sigma' \prec \sigma, \exists \sigma'' \in \Sigma^*: \sigma' \cdot \sigma^{''} \not\in \varphi.
				\end{array}
				\]
			\end{itemize}

The terms impartiality and anticipation are introduced as requirements of monitors in other works related to runtime verification  \cite{DBLP:conf/setss/Leucker16}.

\subsection{Programs and properties in the monolithic and distributed cases}
\label{prelim:mono:dist}
We consider monitoring programs with deterministic behavior, where
in every execution of the program, it consumes an input, and emits an output.
Let $I$ denote a finite set of inputs and $O$ denote a finite set of outputs.
The alphabet $\Sigma = I \times O$, and $\Sigma^*$ is the set of finite words over $\Sigma$.

Since we focus on deterministic programs, for any given alphabet $\Sigma = I \times O$, we consider $\ws \subseteq \Sigma^*$ where  $\ws$ is the set of all finite words over alphabet $\Sigma$ that do not contain input-output events which have the same input values but differ in their output values.\footnote{$\Sigma^\omega$ denotes all infinite words over alphabet $\Sigma$ satisfying this determinism condition.}
We consider $\ws \subseteq \Sigma^*$, where $\forall \sigma \in \Sigma^{\#}$, the following condition holds:
\vspace{-0.5em}
\[
\forall i \in [1, |\sigma|], \forall j \in[1, |\sigma|], \text{ if }
(\Pi_1(\sigma_i) = \Pi_1(\sigma_j) \text{ then } \Pi_2(\sigma_i) = \Pi_2(\sigma_j).
\]	

We consider monitoring both inputs and outputs of programs with deterministic behavior.
A single execution of such a program is an input-output event $(i,o) \in I \times O$, where $I$ denote a finite set of inputs, and $O$ denotes a finite set of outputs.

We are interested in checking whether  a program satisfies data minimality properties  (introduced later in Sections \ref{sec:min:mono}, and \ref{sec:dataMin:distributed}). These properties can be modeled as hyper-properties \cite{hyp1,Finkbeiner2015}, where a hyper-property is a set of sets of traces. When we consider monitoring of hyperproperties, we need to consider multiple executions of the program being monitored, and analysis of sets of traces \cite{Bonakdarpour2016}.

Consider the program to be some service provided by a web server. In practice, such a (service) functionality is not just used once, but multiple times with (different) inputs (e.g., different clients invoking the service).
Thus, when we consider observing (monitoring) input-output behavior of such a program at runtime, we actually observe several executions of the program (i.e, can be considered as monitoring the program executed repeatedly in a loop).

We thus can consider the program in-loop as the program being monitored, 
and data minimality properties that we consider in this work can be formalized as normal trace properties, and the monitoring problem can be reduced to analysis of a single trace.

\begin{remark}[Independence of events]
Note that in the program in-loop each input-output computation is independent. The output produced by the program in each execution is only dependent on the input consumed in that particular execution.
\end{remark}

\subsubsection{Monolithic case.}
In the monolithic case, the program has a single input source.
We denote a deterministic  program in the monolithic case  as  ${\calF}:I \rightarrow O$, where $I$ denote a finite set of inputs, and $O$ denotes a finite set of outputs.
The language of $\calF$ is denoted as $\calL({\calF})$, where $\calL( {\calF}) = \{ (i,o) \in I \times O: o = \calF(i)\}$, and $\calL( {\calF}) \subseteq I \times O$.
Note that $\forall i \in I, \forall o \in O, (i,o) \in \calL( {\calF}) \implies (\forall o' \neq o \in O: (i,o') \not\in  \calL(\calF)  )$.

We consider monitoring both inputs and outputs of a program ${\calF}$.
A single execution of ${\calF}$ is an input-output event $(i,o) \in I \times O$.
We consider observing (monitoring) the input-output behavior over several executions of the program ${\calF}$.

Let program ${\calP}$ denote ${\calF}$ executed repeatedly in a loop.
An execution of  ${\calP}$  is an infinite sequence of input-output events $\sigma \in \Sigma^{\omega}$, where $\Sigma = I \times O$.
The \emph{behavior} of program ${\calP}$ is denoted as $\exec({\calP}) \subseteq \Sigma^{\omega}$.
The \emph{language} of ${\calP}$ is denoted by $\calL({\calP})$ = $\{\sigma \in \ws | \exists \sigma' \in \exec( {\calP}) \wedge \sigma \pref \sigma'\}$ i.e. $\calL( {\calP})$ is the set of all finite prefixes of the sequences in $\exec({\calP})$.
Note that $\calL( {\calP})$ is prefix-closed, i.e., prefixes of any word that belongs to  $\calL( {\calP})$ also belong to $\calL( {\calP})$.

\paragraph{Properties.}
A property $\varphi$ over a finite alphabet $\Sigma$ defines a set $\varphi\subseteq \ws$.
A program $\calP \models \varphi$ iff $\calL(\calP) \subseteq  \calL(\varphi)$.
Given a word $\sigma \in \ws$, $\sigma \models \varphi$ iff $\sigma\in\calL(\varphi)$.

	\begin{figure}[t]
		\centering
		\hspace{-1em}
		\subfloat[Example of a deterministic monolithic program $\calF$\label{fig:func:mono}]{{\includegraphics[scale=0.55]{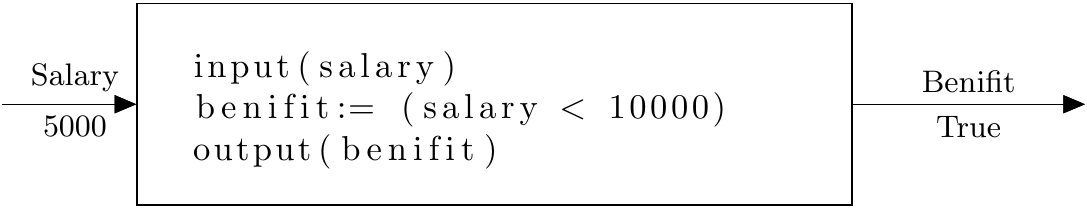} }}%
		\quad
		\subfloat[Example of $\calP$ (repeated execution of $\calF$)\label{fig:prog:mono}]{{\includegraphics[scale=0.50]{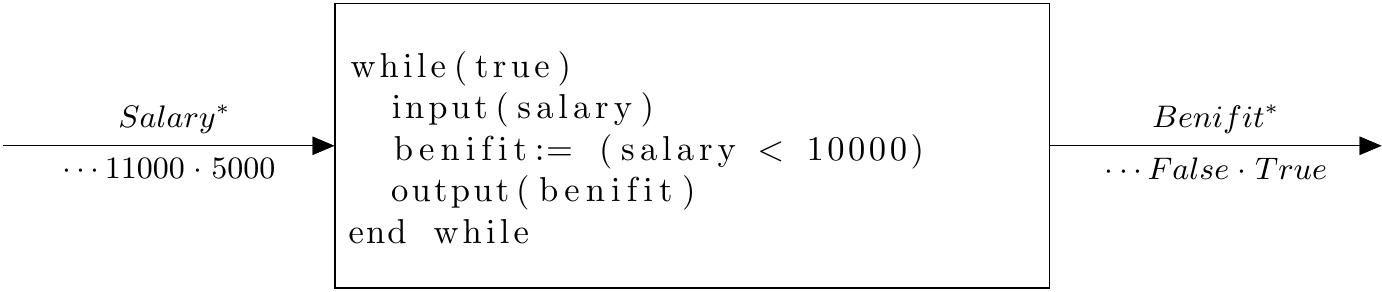} }}%
		\caption{Example of a monolithic program $\calF$ and its associated program-in-loop $\calP$}
	\label{fig:mono:example}
	\end{figure}
%
\begin{example}[Deterministic monolithic program $\calF$ and its corresponding program-in-loop $\calP$]	
Let us consider a simple example illustrated in Figure \ref{fig:mono:example}. 
An example program $\calF: I \rightarrow O$ is illustrated in Figure \ref{fig:func:mono}, that takes salary information (which is an integer, i.e., set of possible inputs $I = \bbn$), and returns whether eligible for benefits or not (i.e., the set of possible outputs $O = \bbb$).
The output of the program is $\true$ if salary is less than 10000, and $\false$ otherwise.

Figure \ref{fig:prog:mono} illustrates an example of program-in-loop $\calP$ corresponding to repeated execution of $\calF$ in Figure  \ref{fig:func:mono}.
Thus, input to $\calP$ is a sequence of inputs events $\sigma_I \in I^*$, and the output is a sequence of outputs $ \sigma_O \in O^*$.
In this example, $\sigma_I = 5000 \cdot 11000 \cdots$, and $\sigma_O = \true \cdot \false \cdots$.
The set of input-output events $\Sigma = \bbn\times \bbb$,  and
a finite prefix of an execution of $\calP$ is $(5000, \true)\cdot(11000, \false)$,  where in the first iteration of the while-loop, input is 5000 and output is $\true$, and in the second iteration, input is 11000  and  output is $\false$.
 $(5000, \true)\in \calL(\calP)$,  and   $(5000, \true)\cdot(11000, \false)\in \calL(\calP)$.
\end{example}	
%
\subsubsection{Distributed case.}
In the distributed case, we consider that the deterministic program has  more than one input sources. In every execution, it consumes an input from each of its source and it emits an output.
We consider a finite number of input sources $n \geq 1$,
where the set of input events $I = I_1\times\cdots\times I_n$ where for all $i \in [1,n] $, $I_i$ is a finite set of possible inputs for input source $i$, and an input event $(i_1,\cdots,i_n)\in I$, where  $i_j\in{I_{j}}$.
In every execution of the program, it consumes an input event $(i_1,\cdots,i_n)\in {I}$, and emits an output event $o\in {O}$, where ${O}$ is a finite set of possible outputs.
A deterministic program in the distributed case is denoted as ${\calDF}: {I_1} \times \cdots \times {I_n}$ $\rightarrow$ ${O}$.

Similar to the monolithic case, we consider monitoring inputs and outputs of $\calDF$.
Let program ${\calDP}$ denote program ${\calDF}$ executed repeatedly in a loop, and let $\Sigma = I \times O$ with ${I} = {I_1}\times\cdots\times {I_n}$.
An execution of  ${\calDP}$  is an infinite sequence of input-output events $\sigma \in \Sigma^{\omega}$.
The \emph{behavior} of program ${\calDP}$ is denoted as $\exec({\calDP}) \subseteq \Sigma^{\omega}$.
The \emph{language} of  ${\calDP}$ is denoted by $\calL({\calDP})$ = $\{\sigma \in \ws | \exists \sigma' \in \exec( {\calDP}) \wedge \sigma \pref \sigma'\}$ i.e. $\calL( {\calDP})$ is the set of all finite prefixes of the sequences in $\exec({\calDP})$.
%

	\begin{figure}[t]
	\centering
	\hspace{-1em}
	\subfloat[Example of a program $\calDF$ in the distributed case  $\calF$ \label{fig:func:dist}]{{\includegraphics[scale=0.55]{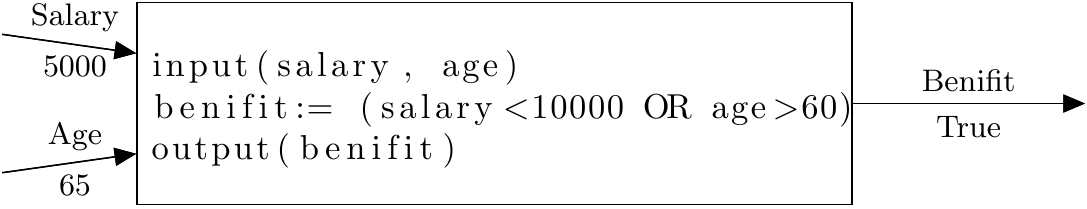} }}%
	\quad
	\subfloat[Example of $\calDP$ (repeated execution of $\calDF$). \label{fig:prog:dist} ]{{\includegraphics[scale=0.5]{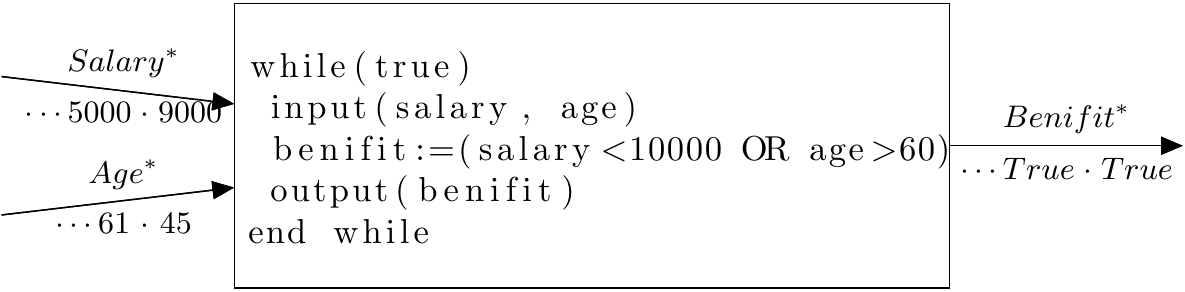} }}%
	\caption{Example of a program $\calDF$ in the distributed case  and its associated program-in-loop $\calDP$}
	\label{fig:dist:example}
\end{figure}

\begin{example}[Program in the distributed case $\calDF$, and its corresponding program -in-loop $\calDP$]
Let us consider a simple example illustrated in Figure \ref{fig:dist:example}.
An example program $\calDF: I_1\times\cdots\times I_n \rightarrow O$ is illustrated in Figure \ref{fig:func:dist}.
In this example, the program has two input sources,  salary which is an integer $I_1 = \bbn$,  and age which is also an integer $I_2 = \bbn$.
It checks eligibility for benefits depending on both salary and age information, and returns a Boolean as output i.e., $O = \bbb$.
The output of the program is $\true$ if salary is less than 10000, or if age is greater than 60.

Figure \ref{fig:prog:dist} illustrates an example of program $\calDP$ which corresponds to repeated execution of $\calDF$ in Figure  \ref{fig:func:dist}.
Thus, input to $\calDP$ is a sequence of inputs events $\sigma_I \in I^*$ where $I = I_1 \times \cdots \times I_n$, and the output is a sequence of outputs
$ \sigma_O \in O^*$.
In this example, we have two input sources (salary and age), and example input word is $\sigma_I = (9000, 45) \cdot (5000, 61) \cdots$, and $\sigma_O = \true \cdot \true \cdots$.
The set of input-output events $\Sigma = I\times O$,  where $I = \bbn \times \bbn$ and $O = \bbb$, and
a finite prefix of an execution of $\calDP$ is $((9000, 45), \true)\cdot((5000, 61), \true)$,  where in the first iteration of the while-loop, input is $(9000, 45)$ and output is $\true$, and in the second iteration, input is $(5000, 61)$ and  output is $\true$.
$((9000, 45), \true)\in \calL(\calDP)$,  and   $((9000, 45), \true)\cdot((5000, 61), \true)\in \calL(\calDP)$.
\end{example}
%

\section{Monolithic case: Data minimality and detection of (non) minimality via monitoring}
\label{sec:monolithic}
In this section, we will focus on the runtime monitoring framework for the monolithic case.
We first recall and formally introduce the data minimality principle\footnote{Our presentation of minimisation here is slightly different from the one given in \cite{ASS17dm}. The main differences are that we define data minimality as a derived concept from strong dependency instead of a characterization from the definition of minimisers.}, and  we introduce the notion of \emph{non-minimality} (Section \ref{sec:min:mono}).
Later, we define a monitoring mechanism to detect  minimality (resp. non-minimality) by observing input-output behavior of a program (Section \ref {sec:mono:monitor}).
We show that (non) minimality property is monitorable.
Detection of satisfaction of minimality property is not possible in general via monitoring. 
However,  when the input domain of the program being monitored is bounded and the monitor has knowledge about the input domain, it is possible to check for satisfaction of minimality.

\subsection{Data minimality in the monolithic case}
\label{sec:min:mono}

We first introduce the data minimality principle in the monolithic case.
Data minimality ensures that the range of inputs provided to a program is reduced such that when two inputs result in the same response, then one of them can be considered redundant. Ideally, a program satisfying data minimisation principle should be one such that the cardinality of the output domain is equal to the cardinality of the input domain.

\begin{figure}[t]
	\centering
	{\includegraphics[scale=0.6]{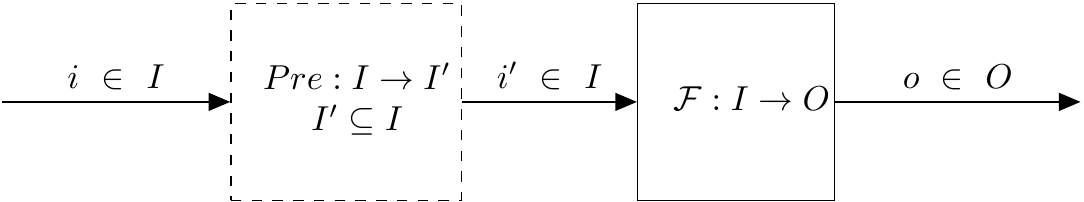}}
	\caption{Input data pre-processor (Monolithic case).}
	\label{fig:mono}
\end{figure}

In Definition~\ref{def:mono-mini2}, as illustrated in Figure \ref{fig:mono}, we assume that there is a data pre-processor which is a function ${{Pre}}$ from $I$ to $I$ that transforms inputs before they are fed to an \emph{un-trusted} program $\calF: I \rightarrow O$.

\begin{definition}[Pre-processor]
\label{def:pre:mono}
Given a program $\calF: I \rightarrow O$, we say that $Pre: I \rightarrow I$ is a pre-processor for $\calF$ iff:
	\begin{enumerate}
		\item $\forall i \in I: \calF(Pre(i)) = \calF(i)$.
		\item $\forall i \in I: Pre(i) = Pre(Pre(i))$.
	\end{enumerate}
\end{definition}
Condition 1 states that the pre-processor should not change the behavior of the program.
For any input $i\in I$, the output that the program produces by consuming the pre-processed input should be equal to the output it produces by directly consuming the input $i$.

Condition 2 states that for any input $i\in I$, if we feed the pre-processed input to the pre-processor again, then it returns back the same pre-processed input.

Pre-processors perform some degree of domain reduction, and 
$\range(Pre) \subseteq I$ (with $\range(Pre)$ denoting the range of function $Pre$, i.e., $\{ Pre(i) | i \in I\}$. 
In case there is no pre-processor, in theory it could be considered that there is a pre-processor that is the identity function. 

\begin{figure}[t]
	\centering
	{\includegraphics[scale=0.85]{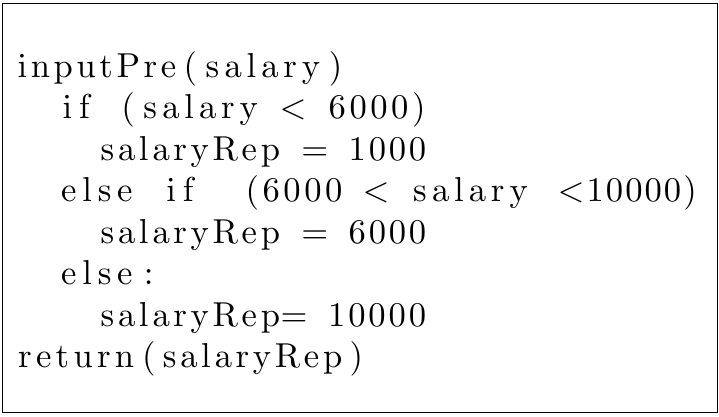}}
	\caption{Example of an input pre-processor for program $\calF$ illustrated in Figure\ref{fig:func:mono}.}
	\label{fig:eg:pre}
\end{figure}

\begin{example}[Input data pre-processor]
Figure \ref{fig:eg:pre} presents an example input pre-processor for the program $\calF$ illustrated in Figure\ref{fig:func:mono}.
If salary is less than 6000 then it is mapped to representative 1000, if salary is greater than 6000 and less than 10000 it is mapped to 6000, and it is mapped to 10000 otherwise.
\end{example}

Definition~\ref{def:mono-mini2} expresses that a program ${\calF}:{I} \rightarrow {O}$, where ${I}$ is the set of possible inputs, $I'\subseteq I$ and ${O}$ is the set of possible outputs, is monolithic minimal for $I'$ if for any two inputs $ i_1, i_2 \in {I'}$, where $i_1$ is different from $i_2$,
the output that program ${\calF}$ produces for input $i_2$ should differ from the output that it produces for input $i_1$.

\begin{definition}[Monolithic minimality of program ${\calF}$]
\label{def:mono-mini2}
A program ${\calF}: {I} \rightarrow {O}$  is monolithic minimal for ${I' \subseteq I}$ \emph{iff} the following condition holds:
\[
\forall i_1, i_2 \in {I'}, \text{ if } i_1 \neq i_2 \text{ then } {\calF}(i_1) \neq {\calF}(i_2).
\]
\end{definition}

\begin{remark}
Note that when we say that $\calF$ is minimal (non-minimal), we  mean that the composition of $\calF$ with a  pre-processor (${{Pre}}: I \rightarrow I'$) is minimal (non-minimal). We consider the scenario illustrated in Figure \ref{fig:eg:pre}.
If composition of $\calF$ with ${{Pre}}$ is minimal (non-minimal), and if ${{Pre}}$ is the identity function then  $\calF$ itself is  minimal (non-minimal).
\end{remark}

We now introduce the notion of \emph{non-minimality} of program ${\calF}$, which is obtained in a straightforward manner by negating the constraint for minimality of program ${\calF}$  in Definition~\ref{def:mono-mini2}.
When a given deterministic program ${\calF}$ satisfies this non-minimality property, then the input given to this program are not minimized in the best possible manner.

Definition~\ref{def:non-mono-min} expresses that a given program ${\calF}:{I} \rightarrow {O}$, where ${I}$ is the set of possible inputs, $I'\subseteq I$ and ${O}$ is the set of possible outputs, ${\calF}$ is monolithic non-minimal for $I'$ if there exists two inputs $ i_1, i_2 \in {I'}$, where $i_1$ is different from $i_2$, and the program ${\calF}$ produces the same output for inputs $i_1$ and $i_2$.
\begin{definition}[Monolithic non-minimality of program ${\calF}$]
	\label{def:non-mono-min}
	A program ${\calF}: I \rightarrow O$  is  non-minimal for $I'\subseteq I$ iff the following condition holds:
	\[
	\exists i_1, i_2 \in I': i_1 \neq i_2 \wedge {\calF}(i_1) = {\calF}(i_2).
	\]
\end{definition}
\begin{example}[Monolithic non-minimality]
	Let us consider the program presented in Figure \ref{fig:func:mono} as $\calF$ , and the function presented in Figure \ref{fig:eg:pre} as the input data pre-processor.
	Thus, $I'$ in this example is $\{1000, 6000, 10000\}$.
	Program $\calF$ is monolithic non-minimal w.r.t $I'$ since there are two elements 1000 and 6000 in $I'$, and  ${\calF}(1000) = {\calF}(6000) = \true$.
\end{example}

We are interested in defining minimality (resp. non-minimality) as trace properties. As discussed in preliminaries, let ${\calP}$ denote a program that executes program  ${\calF}:{I} \rightarrow {O}$ repeatedly.
The language of ${\calP}$ is $\calL(\calP) \subseteq \Sigma^{*}$, where $\Sigma=I \times O$ (see Section \ref{sec:prelim}).

Monolithic minimality property $\varphi_{m} \subseteq \ws$, is the set of all words in  $ws$, such that for any word $\sigma \in \varphi_{m}$,  for any two events at different indexes in $\sigma$, if the projection on inputs of the two events differ, then the  projection on outputs of the two events should also differ. Property $\varphi_{m}$ is formally defined as follows:
\begin{definition}[Monolithic minimality property $\varphi_{m}$]
	\label{def:mono-mini3}
	Given alphabet $\Sigma$, where $\Sigma= I \times O$, property $\varphi_{m} \subseteq \ws$, is the set of all words belonging to $\ws$ satisfying the following constraint:
	\[
	\begin{array}{ll}
	\forall \sigma \in \varphi_m,  \\
	\quad \forall i \in [ 1, |\sigma| ], \forall j \in [1, |\sigma| ], \\
	\quad \quad \text{ if } (\Pi_1(\sigma_i) \neq \Pi_1(\sigma_j)) \text{ then } (\Pi_2(\sigma_i) \neq \Pi_2(\sigma_j))
	\end{array}
	\]
\end{definition}
%
%
\begin{remark}[$\varphi_m$ is prefix-closed]
	Note that all prefixes of all word belonging to $\varphi_m$ also belong to $\varphi_m$, that is, property $\varphi_m$ is prefix-closed.
\end{remark}
A prefix of an execution of a program  $\sigma \in \calL(\calP)$ where  $\calL(\calP) \subseteq \ws$ satisfies property $\varphi_m$ iff $\sigma \in \varphi_{m}$.

\begin{example}[Monolithic minimality property]
Consider program $\calP$ to be the example program illustrated in Figure~\ref{fig:prog:mono}.
Consider a prefix of an execution of this program $\sigma_1 = (5000, \true)\cdot(11000, \false)$ which belongs to $\calL(\calP)$.
We have $\sigma_1 \in \varphi_m$.
Consider another prefix of an execution of this program $\sigma_2 = (5000, \true)\cdot(11000, \false)\cdot (8000, \true)$ where $\sigma_2 \in \calL(\calP)$.
Note that $\sigma_2 \not\in \varphi_m$ since if we consider  input-output events at index 1 and index 3, input values in these two events differ (5000 and 8000) but the output values are equal ($\true$ in both the events).
\end{example}
We now define monolithic non-minimality property, which is negation of the minimality property $\varphi_m$ introduced in Definition~\ref{def:mono-mini3}.
\begin{definition}[Monolithic non-minimality property $\overline{\varphi_{m}}$]
	\label{def:mono-nonmini3}
	Given alphabet $\Sigma$ where $\Sigma= I \times O$, property $\overline{\varphi_{m}}$ is the set of all words in $\ws$ satisfying the following constraint:
	\[
	\begin{array}{ll}
	\forall \sigma  \in \overline{\varphi_{m}}:  \\
	\quad \exists i \in [ 1, |\sigma| ], \exists j \in [1, |\sigma| ]:\\
	\quad \quad  (\Pi_1(\sigma_i) \neq \Pi_1(\sigma_j) \wedge \Pi_2(\sigma_i) = \Pi_2(\sigma_j))
	\end{array}
	\]
\end{definition}
\begin{remark}[Property $\overline\varphi_m$ is extension-closed]
Note, that property $\overline\varphi_m$ is extension closed, i.e., for any word $\sigma$ that belongs to $\varphi_m$, every possible extension of $\sigma$ also belongs to $\overline\varphi_m$.
Formally,  $\forall \sigma \in \ws: \sigma \in \overline\varphi_m \implies(\forall \sigma' \in \ws: \sigma\pref \sigma' \implies \sigma'\in \overline\varphi_m)$.
\end{remark}
\begin{example}
	Consider the example program $\calP$  illustrated in Figure~\ref{fig:prog:mono}.
	Consider a prefix of an execution of this program $\sigma_1 = (5000, \true)\cdot(11000, \false)$ which belongs to $\calL(\calP)$.
	$\sigma_1 \not\in \overline\varphi_m$.
	Consider another prefix of an execution of this program $\sigma_2 = (5000, \true)\cdot(11000, \false)\cdot (8000, \true)$ where $\sigma_2 \in \calL(\calP)$.
	Note that $\sigma_2 \in \overline\varphi_m$ since if we consider  input-output events at index 1 and index 3, input values in these two events differ (5000 and 8000) but the output values are equal ($\true$ in both the events).
	Any possible continuation of $\sigma_2$ also belong to $\overline\varphi_m$.
\end{example}

\begin{lemma}[$\overline{\varphi_{m}} = \ws \setminus \varphi_{m}$]
\label{lem:neg:dm}
Note that $\overline{\varphi_{m}}$ is the negation of property $ \varphi_m$, that is, $\overline{\varphi_{m}} = \ws \setminus \varphi_{m}$.
A word $\sigma \in \ws$ satisfies $\overline{\varphi_{m}}$  if $\sigma\in \overline{\varphi_{m}}$.
It follows that:
\begin{itemize}
\item $ \forall \sigma \in \ws, \sigma \in \varphi_m \implies \sigma \not\in \overline\varphi_m$;
\item $  \forall \sigma \in \ws, \sigma \in \overline\varphi_m \implies \sigma \not\in \varphi_m$.
\end{itemize}
\end{lemma}
Lemma \ref{lem:neg:dm} is immediate consequence of Definitions \ref{def:mono-mini3} and \ref{def:mono-nonmini3}.

Let us consider $\calF$ and its corresponding program-in-loop $\calP$.
The following theorem states that,
if there exists an observation of an execution of program $\calP$ that satisfies the non-minimality property, then program $\calF$ is non-minimal.
If every word that belongs to $\calL(\calP)$ also belongs to property $\varphi_m$ (i.e., every possible observation of execution of $\calP$ satisfies $\varphi_m$), then $\calF$ is monolithic minimal. 
\begin{theorem}
Given $\calF: I \rightarrow O$, let $\calP \subseteq \ws$ where $\Sigma=I \times O$ correspond to the program-in-loop for  $\calF$, the following properties hold:	
\begin{itemize}
	\item $\calF$ is monolithic non-minimal iff $\exists \sigma\in \ws: \sigma \in \calL(\calP) \wedge \sigma \in \overline\varphi_m $;
	\item $\calF$ is monolithic minimal iff $\forall \sigma \in \ws: \sigma \in \calL(\calP) \implies \sigma \in \varphi_m$. 	
\end{itemize} 	
\end{theorem}
\subsection{Monitoring mechanism to detect (non) minimality}
\label{sec:mono:monitor}
We consider monitoring input-output behavior of program ${\calP}$, where ${\calP}$ is repeated execution of program ${\calF}: I \rightarrow O$.
Inputs to program $\calF$ may be fed by a user (which can be considered as monitoring after deployment), or by a $testInputGenerator$ which can be considered as testing  ${\calF}$ for data minimality in a controlled environment prior to deployment.

By monitoring  ${\calP}$ (input-output behavior of several executions of program ${\calF}$), we are interested in checking whether an  execution of ${\calP}$ satisfies (resp. violates) minimality property $\varphi_m$.
\begin{figure}[t]
	\centering
	{\includegraphics[scale=0.7]{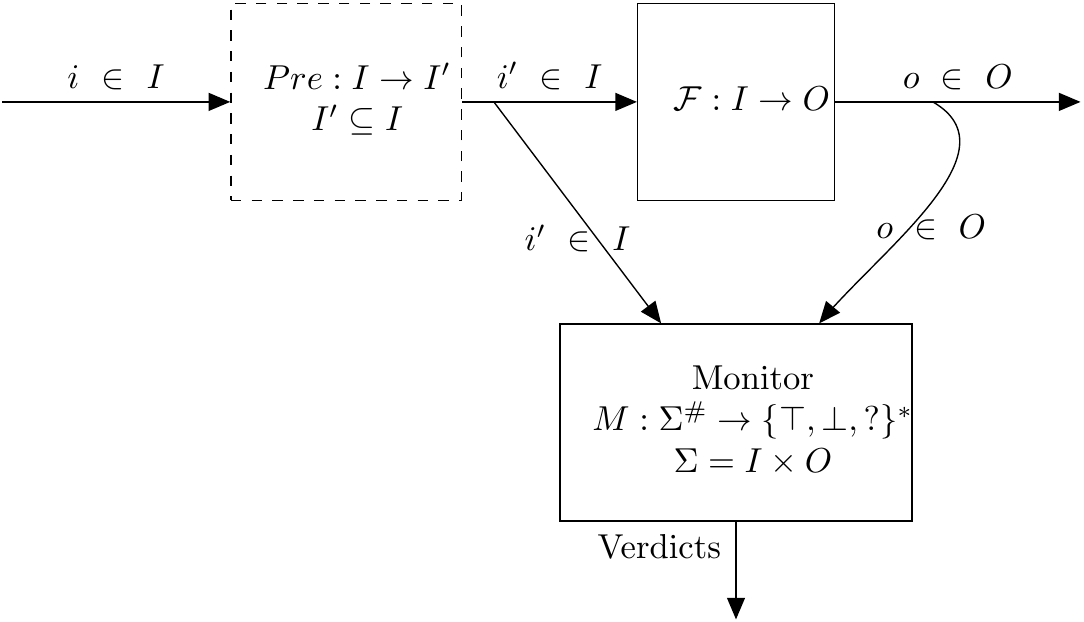}}
	\caption{Problem overview: monolithic case.}
	\label{fig:overview}
\end{figure}

The general context of the proposed monitoring approach is depicted in Figure~\ref{fig:overview}.
The inputs that the user (or $testInputGenerator$) provides belong to the set $I$.
We consider that inputs from the user may be first pre-processed by a data pre-processor ($Pre$), and the pre-processed input that belongs to the set $I'$ (where $I' \subseteq I$) is fed as input to the \emph{untrusted}  program ${\calF}$.
For each execution of ${\calF}$, the monitor observes both the pre-processed input and the output of ${\calF}$.

Note that we also assume that the monitor cannot observe and is not aware of the actual inputs provided by a user.
Moreover, a pre-processor may not exist (i.e., can be considered as the identity function). We also assume that the monitor is unaware whether a pre-processor exists or not, and also does not know about its behavior and the set of all possible outputs $I'$ of the pre-processor.
The monitor observes the pre-processed input $i'$ at runtime, that belongs to $I'$ which also belongs to $I$.

\begin{figure}[t]
	\centering
	{\includegraphics[scale=0.75]{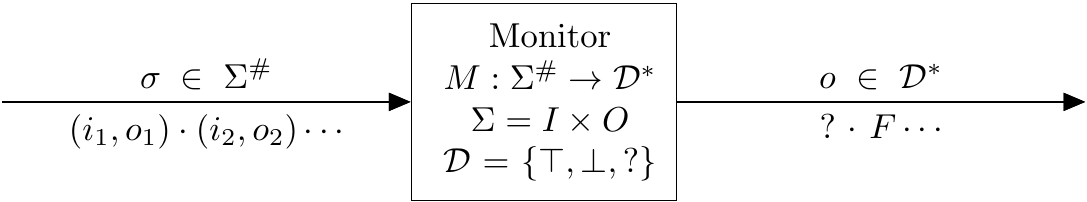}}
	\caption{Monitor $M$.}
	\label{fig:monitor}
\end{figure}

Let us recall that the set of input-output events that the monitor can observe (receive) as input is denoted using $\Sigma$, where $\Sigma = I \times O$, and an input-output event is denoted as $(i, o )$ where $i\in I$ and $o\in O$.
After $n$ executions of the program ${\calF}$, the monitor  observes a word $\sigma = (i_1, o_1), \cdots, (i_n, o_n) \in \calL(\calP)$ as input.
This is illustrated in Figure~\ref{fig:monitor}.

After each execution of $\calF: I \rightarrow O$ (i.e., in every iteration of  program {$\calP$}), the monitor observes the (pre-processed) input and the output of program $\calF$ in that particular iteration (step) of {$\calP$} .

For any word $\sigma\in \ws$ (current observation of execution of {$\calP$}) where $|\sigma| >1$,  $M_{\varphi_m}$ as per Definition \ref{def:rv:mon} is a monitor for property $\varphi_m$.
The monitor returns $\true$ ($\top$) when $\sigma$ followed by any extension of it satisfies the minimality property $\varphi_m$.
The monitor returns $false$ ($\bot$) when the current observation of execution of {$\calP$} followed by any extension of it violates $\varphi_m$ (resp. satisfies $\overline\varphi_m$).
It returns $?$ (unknown) for the current observation if the other two cases do not hold.

We do not define properties in a finite representation (such as automata) or using some logic such as LTL, and moreover properties we consider are not related to checking constraints on the order in which certain actions should happen.
Our properties $\varphi_m$ (resp. $\overline{\varphi_m}$) are related to checking some relation between the input-output values.

We thus reduce checking whether the property is satisfied (resp. violated) for every extension of the current observation,  to checking whether the current observation satisfies (resp. violates) some conditions.

We now introduce  function $\satMin$ that is defined based on definitions of properties $\varphi_m$ (resp. $\overline{\varphi_m}$).
This function is used to check whether the current observation $\sigma$ satisfies property $\varphi_m$ (resp. $\overline{\varphi_m}$) .

Function $\satMin: \ws \rightarrow \bbb$, takes an input-output word and returns a Boolean as output. It is defined as follows:
\[
\begin{array}{lll}
\satMin(\sigma)& =
\begin{cases}
\true & \mbox{if}\
\begin{array}{ll}
\forall i \in [ 1, |\sigma| ], \forall j \neq i \in [1, |\sigma| ]: \\
\quad\quad
\Pi_1(\sigma_i) \neq \Pi_1(\sigma_j) \implies \Pi_2(\sigma_i) \neq \Pi_2(\sigma_j)
\end{array}
\\
\false & Otherwise\\
\end{cases}
\end{array}
\]
For any given word $\sigma$, $\satMin(\sigma)$ returns $\true$ if $\sigma \in \varphi_m$, and returns $\false$ otherwise, i.e, if it returns $\false$, then $\sigma \in \overline{\varphi_m}$.

Note that from the definition of $\overline{\varphi_m}$, 
if $\sigma \in \overline{\varphi_m}$, then $\exists i \in [ 1, |\sigma| ], \exists j \neq i \in [1, |\sigma| ]: (\Pi_1(\sigma_i) \neq \Pi_1(\sigma_j) \wedge \Pi_2(\sigma_i) = \Pi_2(\sigma_j))$. 

\remove{
Similarly, function  $\satNonMin: \Sigma^* \rightarrow \bbb$, takes an input-output word $\sigma$ and returns a Boolean as output. It is defined as follows:

\[
\begin{array}{lll}
\satNonMin(\sigma)& =
\begin{cases}
\true & \mbox{if}\
\begin{array}{ll}
 \exists i \in [ 1, |\sigma| ], \exists j \neq i \in [1, |\sigma| ]: \\
\quad \quad (\Pi_1(\sigma_i) \neq \Pi_1(\sigma_j) \wedge \Pi_2(\sigma_i) = \Pi_2(\sigma_j))
 \end{array}
 \\
\false & Otherwise\\
\end{cases}
\end{array}
\]
For any given word $\sigma$, $\satNonMin(\sigma)$ returns $\true$ if $\sigma \in \overline\varphi_m$, and returns $\false$ otherwise.
}
The following proposition expresses that if a word $\sigma$ belongs to the property $\overline{\varphi_m}$, then every possible extension of it also belongs to the property $\overline{\varphi_m}$.
\begin{proposition}
	\label{prop1}	
	$\forall \sigma \in \ws, \text{ if } \sigma\in \overline{\varphi_{m}} \text{ then }(\forall \sigma'\in \ws: \sigma\pref \sigma' \implies \sigma'  \in \overline{\varphi_{m}})$.
\end{proposition}

\begin{example}
	Consider the example program $\calP$  illustrated in Figure~\ref{fig:prog:mono}.
	Consider a prefix of an execution of this program $\sigma_1 = (5000, \true)\cdot(11000, \false)$ which belongs to $\calL(\calP)$.
	We have $\satMin(\sigma_1) = \true$.
	Consider another prefix of an execution of this program $\sigma_2 = (5000, \true)\cdot(11000, \false)\cdot (8000, \true)$ where $\sigma_2 \in \calL(\calP)$.
	We have $\satMin(\sigma_2) = \false$.
	For any word $\sigma' \in \ws$, $\satMin(\sigma_2\cdot \sigma')$ will be $\false$ (i.e., $\sigma_2\cdot \sigma' \in \overline{\varphi_m}$).
\end{example}
\begin{remark}[Condition for conclusive verdict $\bot$]
From Proposition \ref{prop1}, we can reduce the condition of the second case in $M_{\varphi_m}$ as per Definition \ref{def:rv:mon} for property ${\varphi_m}$  to checking whether the current observation $\sigma$ satisfies $\overline\varphi_m$. 
\end{remark}
%
\begin{remark}[Impossibility of checking condition of the first case (satisfaction of property $\varphi_m$)]
\label{prop:undecidable}
Note that regarding the condition of the $\top$ case (satisfaction of $\varphi_m$), checking whether the current observed word $\sigma$ belongs to $\varphi_m$ (i.e.,whether $\satMin(\sigma)$ is $\true$) is not sufficient, and does not ensure that every extension of $\sigma$ will also belong to $\varphi_m$ if $\sigma$ belongs to $\varphi_m$. Thus, testing condition of the first case is not possible in general.
\end{remark}

By providing the monitor with knowledge about the input domain and when the input domain is bounded, it is possible to test the condition of the first case related to satisfaction of property $\varphi_m$.

We also introduce function $\inpExh: I \times \ws \rightarrow \bbb$ where $\Sigma = I\times O$, that is used to test whether every input  belonging to the set $I$ appear at least once in a given input-output word $\sigma \in \ws$. It is defined as follows:
\[
\begin{array}{lll}
\inpExh(I, \sigma)& =
\begin{cases}
\true & \mbox{if}\
\begin{array}{ll}
(\forall i \in I, \exists id \in [1, |\sigma|]: \Pi_1(\sigma_{id}) = i)
\end{array}
\\
\false & Otherwise\\
\end{cases}
\end{array}
\]
%
\begin{remark}[Condition for conclusive verdict $\top$ (satisfaction of property $\varphi_m$)]
	Note that the condition of the first case in $M_{\varphi_m}$ as per Definition \ref{def:rv:mon} for property ${\varphi_m}$ reduces to checking whether $\sigma$ satisfies the following two conditions:
	\begin{itemize}
		\item every input belonging the set of inputs $I$ appear at least once in $\sigma$, i.e., $\inpExh(I, \sigma) = \true$ and
		\item  $\sigma$ satisfies $\varphi_m$, i.e., $\satMin(\sigma) = \true$.
	\end{itemize}
Note that if $\sigma$ contains every input belonging to the set of inputs $I$, and if $\sigma$ satisfies $\varphi_m$, then every possible extension of $\sigma$ also satisfies $\varphi_m$.
\end{remark}
\begin{proposition}
	\label{prop:true}
	Given any word $\sigma \in \ws$, where $\Sigma = I \times O$, and  $|\sigma| >1$, we have;
	\[ \text{ if } (\inpExh(I, \sigma) \wedge \satMin(\sigma) )  \text{ then } (\forall \sigma',   \text{ if } \sigma \pref \sigma' \text{ then }  \sigma' \in {\varphi_m}).\]
	
\end{proposition}

Thus, using Propositions \ref{prop1}, \ref{prop:true}, the conditions of the first two cases in $M_{\varphi_m}$ can be simplified. We present the alternative simplified definition below, where the conditions of the first two cases are reduced to checking whether the observed input word satisfies some conditions.
 
 Consider property $\varphi_m \subseteq \ws$ where $\Sigma= I \times O$.
  Let $\sigma \in \ws$ denote a finite input-output word over the alphabet $\Sigma = I \times O$.
 A monitor for  property $\varphi_m$ (resp. $\overline{\varphi_m}$) is a function $M_{\varphi_m}: \ws \rightarrow \D$, where $D = \{ \top, \bot, ? \}$.  
 For $\sigma = \epsilon$ and any word $\sigma$ of length 1, $M(\sigma) = ?$.
Monitor $M_{\varphi_m}$ is defined as follows:
\begin{definition}[Monitor $M_{\varphi_m}$]
\label{def:monitor:mono2}
A monitor for  property $\varphi_m$ (resp. $\overline{\varphi_m}$) is a function $M_{\varphi_m}: \ws \rightarrow \D$, where $D = \{ \top, \bot, ? \}$ is defined as follows:
	\[
	\begin{array}{lll}
	M_{\varphi_m}(\sigma) & =
	\begin{cases}
	\top & \mbox{if }\  |\sigma|>1 \wedge \inpExh (I, \sigma) \wedge \satMin(\sigma) \\
	\bot & \mbox{if }\  |\sigma|>1 \wedge \neg\satMin(\sigma) \\
	? & Otherwise
	\end{cases}
	\end{array}
	\]
\end{definition}
\begin{proposition}
	\label{prop:monitor:mono}
	 $M_{\varphi_m}$ in Definition \ref{def:monitor:mono2} is a monitor for property $\varphi_m$  (i.e., $M_{\varphi_m}$ satisfies \ref{eq:snd} and \ref{eq:opt}).
\end{proposition}
\begin{table}[t]
	\centering
	\begin{tabular}{|c|c|}
		\hline
		\multicolumn{1}{|c|}{$\sigma$} & \multicolumn{1}{c|}{$M(\sigma)$} \\
		\hline
		{$(5000,\true)$} & $?$  \\
		\hline
		$(5000,\true) \cdot (11000,\false)$ & $?$ \\
		\hline
		$\bm{(5000,\true)} \cdot (11000,\false) \cdot \bm{(8000, \true)}$ & $\bot$ \\
		\hline
		$\bm{(5000,\true) }\cdot (11000,\false) \cdot\bm{(8000, \true)} \cdot (12000, \false)$ & $\bot$ \\
		\hline
$\bm{(5000,\true)} \cdot (11000,\false) \cdot \bm{(8000, \true)}\cdot (12000, \false) \cdots$ & $\bot$ \\
\hline
	\end{tabular}
	\caption{Example illustrating behavior of the monitor $M_{\varphi_m}$.}
	\label{tableExample1}
\end{table}

\begin{example}[Example illustrating behavior of the monitor $M_{\varphi_m}$]

Let us again consider the example program $\calP$  illustrated in Figure~\ref{fig:prog:mono}.
In Table \ref{tableExample1}, we present some example observations of an execution of program $\calP$ being monitored denoted as $\sigma$, and the verdict provided by the monitor for $\sigma$.
Initially, when the first event observed in $(5000, \true)$, the monitor returns verdict unknown (?). In each step current observation is extended with a new event. Let the new event observed in the second step be $(11000, \false)$. For current observation $\sigma = (5000, \true) \cdot (11000, \false)$, the monitor returns verdict unknown.
After observing the third event $(8000, \false)$, the monitor returns verdict $\false$ ($\bot$) for $\sigma = (5000, \true) \cdot (11000, \false) \cdot (8000, \false)$.
\end{example}
\begin{remark}[Monitor with two cases]
Note that when it is not possible to test whether current observation $\sigma$ covers all inputs (i.e., when the input domain $I$ is unknown), it is not possible to compute $\inpExh(I,\sigma)$. In this case, one can consider monitor with two cases (where the first case is merged with the unknown case).
The monitor returns $\bot$ indicating violation of minimality (resp. satisfaction of non-minimality), if $\neg\satMin(\sigma)$ and $?$ otherwise.
\end{remark}

\section{Distributed case: Data minimality and detection of (non) minimality via monitoring}
\label{sec:distributed}
In Section~\ref{sec:monolithic}, we considered that the program has a single input source. However, in several domains such as web services, a service provider requires data  from multiple sources.

In this section, we  introduce data minimality (and non-minimality) policies for the distributed case (Section \ref{sec:dataMin:distributed}).
Distributed minimality (resp. distributed non-minimality) are not monitorable in general.
Results we obtained for minimality (resp. non-minimality) in the monolithic case can be extended for {\em strong} distributed minimality (resp. \emph{strong} distributed non-minimality).
We present a monitoring mechanism to detect strong distributed minimality (resp. non-minimality) by observing input-output behavior of a program (Section \ref{sec:mon:dist}).
\begin{figure}[t]
	\centering
			{\includegraphics[scale=0.7]{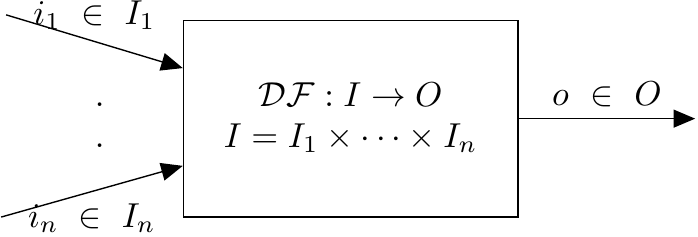}}
	\caption{Program in the distributed case with multiple inputs.}
	\label{fig:distributed}
\end{figure}

A program in the distributed case with multiple input sources is illustrated in Figure~\ref{fig:distributed}.
We consider deterministic programs that can be considered as functions with multiple input sources (e.g., multiple clients).
In each execution of the program, it consumes an input event from all its input sources and it emits an output event.

As discussed in Section~\ref{sec:prelim}, a program in the distributed case with $n$ input sources can be considered as a function,  denoted as ${\calDF}: {I_1} \times \cdots \times {I_n}$ $\rightarrow$ ${O}$.
We consider monitoring input-output behavior of multiple executions of program  ${\calDF}$. Repeated execution of ${\calDF}$ is denoted as ${\calDP}$, where ${\calDP}: I^* \rightarrow O^*$, with $I= {I_1} \times \cdots \times {I_n}$.
In Figure \ref{fig:dist:example}, we present an example of a program $\calDF$ with two input sources and its corresponding program $\calDP$.

Monolithic minimality will be too restrictive in the distributed case where we have multiple input sources. 
Let us consider the following examples taken from \cite{ASS17dm}. Let ${\calDF}$ be the program $\textrm{XOR}: \bbb \times \bbb \rightarrow \bbb$ that takes two Boolean inputs and returns a Boolean as output. Since $\textrm{XOR}(0,0) = \textrm{XOR}(1,1)$, it follows that $\textrm{XOR}$ is not monolithic minimal.  
The program $\textrm{OR}: \bbb \times \bbb \rightarrow \bbb$ is also not monolithic minimal since $\textrm{OR}(0,1) = \textrm{OR}(1,0)$.
Monolithic minimality is in general not suitable (too strong a notion) for programs with multiple input sources.

Let us first understand the notions of distributed minimality, where the program being monitored has multiple input sources.
\subsection{Data minimality in the distributed setting}
\label{sec:dataMin:distributed}
\begin{figure}[t]
	\centering
	\hspace{-1em}
	\subfloat[Single pre-processor \label{fig:distSinglePre}]{{\includegraphics[scale=0.6]{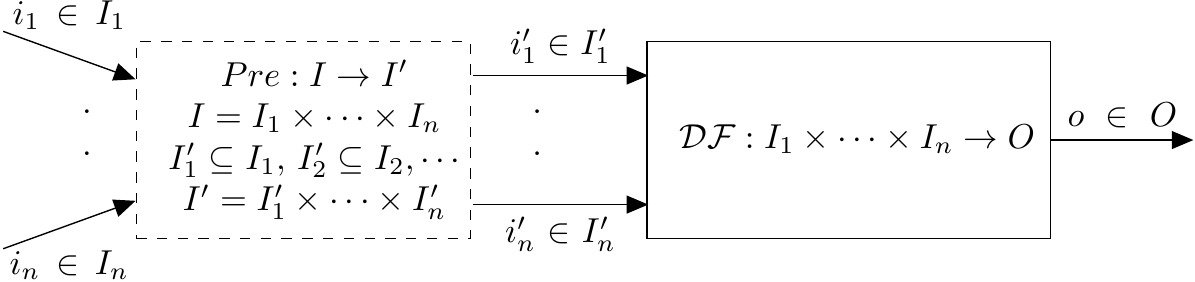} }}%
	\quad
	\subfloat[Multiple pre-processors (one per input source))\label{fig:distMultiPre}]{{\includegraphics[scale=0.6]{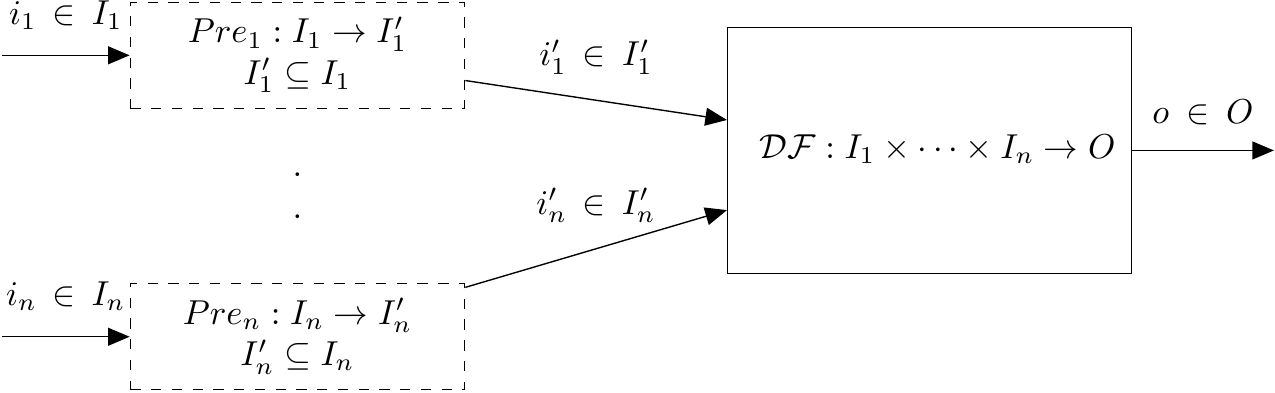} }}%
\caption{Input data pre-processor (Distributed case).}
\label{fig:dist1}
\end{figure}

In the following definitions, similar to the monolithic case, as illustrated in Figure \ref{fig:distSinglePre}, we assume that there may be a pre-processor which can be considered as a function $Pre: I_1 \times \cdots \times I_n \rightarrow I'_1 \times \cdots \times I'_n$ (where for all $i \in [1,n]$, $I'_i \subseteq I_i$) that transforms user inputs before they are fed to the program.
As illustrated in Figure \ref{fig:distMultiPre}, there can be multiple pre-processors one for each input source.

Pre-processor(s) for which the cardinality of the output domain is lesser than the cardinality of the input domain may not exist, and in that case, they 
can be considered as  identity function(s) (forwarding user inputs to the program).

Definition \ref{def:pre:mono} of a pre-processor in the monolithic case can be extended to the distributed case, and we omit details here (See \cite{ASS17dm} for details and definitions\footnote{In \cite{ASS17dm}, pre-processors are called {\em minimisers}, and minimisers are defined as {\em best minimisers}.}).

We first present the notion of distributed minimality considered in \cite{ASS17dm}, and introduce distributed non-minimality.

\begin{definition}[Distributed minimality of program $\calDF$]
	\label{def:weak:dist:min}
	Program ${\calDF}: I_1\times\cdots\times I_n \rightarrow O$ is  {\em distributed minimal} for $I'\subseteq I$ iff for every input source $I_{id}$ where $id \in \{1,\cdots,n\}$, for any two different values $u,v\in I_{id}$, there are at least two input events $(i_1, \cdots, i_n), (i'_1, \cdots, i'_n) \in I$ which differ in exactly one input source value  (where $i_{id} = u$ and $i'_{id} = v$), and the program ${\calDF}$ produces different output for  $(i_1, \cdots, i_n)$ and $(i'_1, \cdots, i'_n)$. Formally,
	
	\[
	\begin{array}{ll}
	\forall id\in [1,n], \forall u,v \in I_{id} \text{ such that } u\neq v,\\
	\quad\quad\exists i_1, i_2 \in I':((\Pi_{id}(i_1) = u \wedge \Pi_{id}(i_2) = v) \\
	\quad\quad\quad\quad\wedge (\forall j \in [1,n]: j\neq id \implies \Pi_{j}(i_1) = \Pi_{j}(i_2))\\
	\quad \quad\quad \quad \quad\quad \quad \quad \wedge {\calDF}(i_1) \neq {\calDF}(i_2))
	\end{array}
	\]
\end{definition}

\begin{example}
Distributed minimality is a weakening of the monolithic minimality (Definition \ref{def:mono-mini2}).
For example, the $\textrm{OR}: \bbb\times \bbb \rightarrow \bbb$ function, which was shown not to be  monolithic minimal, is distributed-minimal.
We have two input sources $I_1 \times I_2$ ($\bbb\times \bbb$).
For the first input source, for each possible pair of distinct values in that position (that is $(0,\_)$ and $(1,\_)$), we can find satisfactory input tuples yielding different results (e.g., $((0,0),(1,0))$ since $\textrm{OR}(0,0) \neq \textrm{OR}(1,0)$).\footnote{Note that the pair  $((0,1),(1,1))$ would not satisfy the definition, but this is fine as the definition only requires that at least one such tuple exists.}
Similarly, for input source 2, for each possible pair of distinct values in that position ($(\_,0)$ and $(\_,1)$), we have that the tuples $(0,0)$ and $(0,1)$ satisfy the definition ($\textrm{OR}(0,0) \neq \textrm{OR}(1,0)$).
\end{example}

We now introduce  distributed non-minimality of a program $\calDF$, 
as a negation of Definition \ref{def:weak:dist:min}.

\begin{definition}[Distributed non-minimality]
	\label{def:weak:dist:non:min}
	Program ${\calDF}: I_1\times\cdots\times I_n \rightarrow O$ is {\em distributed non-minimal} for $I'\subseteq I$ iff there is an input source $I_{id}$ where $id \in [1,n]$, such that there exist two different values $u,v\in I_{id}$
	such that for any two input events $i_1$ and $i_2$ that  belong to $I'$ where the value corresponding to input source $id$ is $v$ is one and $u$ in the other,  and the values of other input sources are equal in both $i_1$ and $i_2$, the program produces the same output for $i_1$ and $i_2$. Formally,
	\[
	\begin{array}{ll}
	\exists id\in [1,n], \exists u,v \in I_{id} \text{ such that } u\neq v,\\
	\quad\quad\forall i_1, i_2 \in I':((\Pi_{id}(i_1) = u \wedge \Pi_{id}(i_2) = v) \wedge\\
	\quad\quad\quad\quad(\forall j \in [1,n]: j\neq id \implies \Pi_{j}(i_1) = \Pi_{j}(i_2)))\\
	\quad \quad\quad \quad \quad\quad \quad \quad \implies {\calDF}(i_1) = {\calDF}(i_2)
	\end{array}
	\]
\end{definition}

\begin{remark}[Non-monitoribality of distributed minimality (resp. non-minimality)]
		Both satisfaction and violation of distributed minimality are not monitorable in general.  
		When monitoring a program $\calDP$ (repeated execution of program $\calDF$), 
		detection of violation of distributed minimality also requires all the input domains to be known and bounded, and checking whether the current observation of execution of $\calDP$ covers all the inputs.
\end{remark}

We thus consider a variant of distributed minimality, called as \emph{strong} distributed minimality.
Later in Section \ref{sec:mon:dist}, we show that the results related to monitoring for minimality (respectively non-minimality) in the monolithic case can be extended to monitoring for strong distributed minimality (respectively strong distributed non-minimality) in the distributed case.

\begin{definition}[Strong distributed minimality of program ${\calDF}$]
\label{def:dist:min}
Program ${\calDF}: I_1\times \cdots\times I_n \rightarrow O$,
where for all input sources $id\in [1,n]$, $I_{id}$ is the set of possible inputs from source $id$, $I' \subseteq I$, and $O$ is the set of possible outputs, ${\calDF}$ is {\em strongly distributed minimal} for $I'$ iff:\\
for any two input events $(i_1, \cdots, i_n)$ and $(i'_1, \cdots, i'_n)$ belonging to $I'$ that differ exactly in one element, the output that the program ${\calDF}$ produces for input $(i'_1, \cdots, i'_n)$ is different from the output that it produces for input $(i_1, \cdots, i_n)$. Formally,
\[
\begin{array}{ll}
\forall (i_1, \cdots, i_n), (i'_1, \cdots, i'_n) \in I': \\
\quad(\exists j \in [1,n]: i_{j} \neq i'_{j} \wedge \forall k\in [1,n]: k\neq j \implies i_{k}= i'_{k}) \implies  \\
\quad\quad {\calDF}((i_1, \cdots, i_n)) \neq {\calDF}((i'_1, \cdots, i'_n))
\end{array}
\]
\end{definition}
\begin{remark}
When the number of input sources is one,  strong distributed minimality (Definition \ref{def:dist:min}) reduces to the monolithic minimality (Definition \ref{def:mono-mini2}).
\end{remark}

\begin{example}
Strong distributed minimality is also a weakening of the monolithic minimality (Definition \ref{def:mono-mini2}).
For example, we already saw that the $\textrm{XOR}: \bbb \times \bbb \rightarrow \bbb$ function is not monolithic minimal, since $\textrm{XOR}(0,0) = \textrm{XOR}(1,1)$.
We can easily notice that the $\textrm{XOR}$ function is strong distributed minimal since the output differs for every possible pair of input tuples differing exactly at one position (e.g., $\textrm{XOR}(0,0) \neq \textrm{XOR}(0,1)$).
Distributed minimality is weaker than strong distributed minimality.
We already showed previously that the  $\textrm{OR}$ function is distributed minimal. However it is not strong distributed minimal since input events $(0,1)$ and $(1,1)$ differ at exactly one position and $\textrm{OR}(0,1) = \textrm{OR}(1,1)$.
\end{example}

\begin{definition}[Strong distributed non-minimality of program $\calDF$]
	\label{def:dist:non:min}
	Program ${\calDF: I \rightarrow O}$ is {\em strong distributed non-minimal} for $I' \subseteq I$ iff there exists two input events  $(i_1, \cdots, i_n)$ and $(i'_1, \cdots, i'_n)$ belonging to $I'$ that differ exactly in one element,
	and the output that the program ${\calDF}$ produces for $(i'_1, \cdots, i'_n)$  is equal to the output that it produces for $(i_1, \cdots, i_n)$. Formally,
	\[
	\begin{array}{ll}
	\exists (i_1, \cdots, i_n), (i'_1, \cdots, i'_n) \in I': \\
	\quad(\exists j \in [1,n]: i_{j} \neq i'_{j} \wedge \forall k\in  [1,n]: k\neq j \implies i_{k}= i'_{k}) \wedge  \\
	\quad\quad {\calDF}((i_1, \cdots, i_n)) = {\calDP}((i'_1, \cdots, i'_n))
	\end{array}
	\]
\end{definition}
We now introduce strong distributed minimality property denoted as $\varphi_{sdm}$ based on the definition of distributed minimality (Definition \ref{def:dist:min}).

\begin{definition}[Strong distributed minimality property $\varphi_{sdm}$]
	\label{def:dist-miniProp}
	{\em Strong distributed minimality property} $\varphi_{sdm}\subseteq \ws$, where $\Sigma= I \times O$ and $I= I_1\times \cdots\times I_n $ is the set of all words belonging to $\ws$, such that for any word  $\sigma \in \varphi_{sdm}$, for any two input-output events at different indexes in $\sigma$, let the inputs corresponding to the two event be $(i_1, \cdots, i_n)$ and $(i'_1, \cdots, i'_n)$.
	If only one input source value differ in $(i_1, \cdots, i_n)$ and $(i'_1, \cdots, i'_n)$,  then the projection on outputs of the two input-output events  should differ. Formally,
	\[
	\begin{array}{ll}
	\forall \sigma \in \varphi_{sdm},  \\
	\quad \forall i \in [ 1, |\sigma| ], \forall j\neq i \in [1, |\sigma| ], \\
	\quad \quad \text{ let } \Pi_1(\sigma_i)= (i_1, \cdots, i_n), \Pi_2(\sigma_j)= (i'_1, \cdots, i'_n).\\
	\quad \quad \quad (\exists x \in [1,n]: i_{x} \neq i'_{x} \wedge \forall y\in [1,n]: y\neq x \implies i_{y}= i'_{y}) \implies
	 \Pi_2(\sigma_i) \neq \Pi_2(\sigma_j).	
	\end{array}
	\]
\end{definition}
%
\begin{remark}
Note that property $\varphi_{sdm}$ is prefix-closed.	
\end{remark}	
%
\begin{example}
\label{exanple:dist:minimalP}
Consider program $\calDP$ to be the example program illustrated in Figure~\ref{fig:dist:example}.
	Let $\sigma_1 =  ((5000, 45), \true)\cdot((11000, 45), \false)$ be a prefix of an execution of this program which belongs to $\calL(\calDP)$.
	We have $\sigma_1 \in \varphi_{dm}$.
	Consider another prefix of an execution of this program $\sigma_2 = ((5000, 45), \true)\cdot((11000, 45), \false) \cdot ((12000, 45), \false)$ where $\sigma_2 \in \calL(\calP)$.
	Note that $\sigma_2 \not\in \varphi_{sdm}$ since if we consider  input-output events at index 2 and index 3, the projection of inputs in these events are resp. $ (11000, 45)$ and $(12000, 45)$, and only salary information in these two input events differ. The output values of these two events are equal ($\true$ in both the events at index 2 and 3).
\end{example}


We now define strong distributed  non-minimality property, which is negation of the distributed  minimality property $\varphi_{sdm}$ introduced in Definition~\ref{def:dist-miniProp}.
\begin{definition}[Strong distributed non-minimality property $\overline{\varphi_{sdm}}$]
	\label{def:dist-nonminiProp}
	Given alphabet $\Sigma = I \times O$, where $I= I_1\times \cdots\times I_n $, property $\overline\varphi_{sdm} \subseteq \ws$, is the set of all words in $\ws$ satisfying the following constraint:
	\[
	\begin{array}{ll}
	\forall \sigma  \in \overline{\varphi_{sdm}}:  \\
	\quad \exists i \in [ 1, |\sigma| ], \exists j \neq i \in [1, |\sigma| ] , \text{ with } \Pi_1(\sigma_i)= (i_1, \cdots, i_n), \Pi_2(\sigma_j)= (i'_1, \cdots, i'_n) \text{ s.t. }\\
	\quad \quad((\exists x \in [1,n]: i_{x} \neq i'_{x} \wedge \forall y\in  [1,n]: y\neq x \implies i_{y}= i'_{y}) \wedge  \\
	\quad\quad \quad\quad (\Pi_2(\sigma_i) = \Pi_2(\sigma_j))
	\end{array}
	\]
\end{definition}
\begin{remark}
	Note, that property $\overline\varphi_{sdm}$ is extension closed, i.e., for any word $\sigma$ that belongs to $\varphi_{sdm}$, every possible extension of $\sigma$ also belongs to $\overline\varphi_{sdm}$.
	Formally,  $\forall \sigma \in \ws: \sigma \in \overline\varphi_{sdm} \implies(\forall \sigma' \in \ws: \sigma\pref \sigma' \implies \sigma'\in \overline\varphi_{sdm})$.
\end{remark}

\begin{example}
	\label{exanple:dist:nonMinimalP}
		Let us consider the example program $\calDP$ illustrated in Figure~\ref{fig:dist:example}.
	Consider a prefix of an execution of this program $\sigma_1 = ((5000, 45), \true)\cdot((11000, 45), \false) \cdot ((12000, 45), \false)$ where $\sigma_2 \in \calL(\calP)$.
	Note that $\sigma_1 \in \overline\varphi_{sdm}$ since if we consider  input-output events at index 2 and index 3, the projection of inputs in these events are resp. $ (11000, 45)$ and $(12000, 45)$, and only salary information in these two input events differ. The output values of these two events are equal ($\true$ in both the events at index 2 and 3).
	Note that any extension of $\sigma_1$ also belongs to $\overline\varphi_{sdm}$.
\end{example}

\begin{remark}
	Note that $\overline{\varphi_{sdm}}$ is the negation of property $ \varphi_{sdm}$, where $\overline{\varphi_{sdm}} = \ws \setminus \varphi_{sdm}$.
	A word $\sigma \in \ws$ satisfies $\overline{\varphi_{sdm}}$  if $\sigma\in \overline{\varphi_{sdm}}$.
	It follows:
	\begin{itemize}
		\item $ \forall \sigma \in \ws, \sigma \in \varphi_{sdm} \implies \sigma \not\in \overline\varphi_{sdm}$;
		\item $  \forall \sigma \in \ws, \sigma \in \overline\varphi_{sdm} \implies \sigma \not\in \varphi_{sdm}$.
	\end{itemize}
\end{remark}

\begin{theorem}
	Given $\calDF: I \rightarrow O$ where $I =  I_1\times \cdots\times I_n$, let $\calL(\calDP) \subseteq \ws$ with $\Sigma=I \times O$, where $\calDP$ corresponds to the program for  $\calDF$ ($\calDP$ is repeated execution of program $\calDF$).
	The following properties hold:	
	\begin{itemize}
		\item $\calDF$ is strong distributed  non-minimal iff $\; \exists \sigma\in \ws: \sigma \in \calL(\calDP) \wedge \sigma \in \overline\varphi_{dm}$.
		\item $\calDF$ is strong distributed  minimal iff $\; \forall \sigma \in \ws: \sigma \in \calL(\calDP) \implies \sigma \in \varphi_{dm}$.
	\end{itemize} 	
\end{theorem}
\begin{theorem}[Minimality $\implies$ strong distributed minimality $\implies$ distributed minimality]
	\label{th:weak:dist}
	If program  ${\calDF}: I \rightarrow O$ is strong distributed minimal, for $I' \subseteq I$, according to Definition \ref{def:dist:min}, then
	${\calDF}$ is also distributed minimal for $I'$ as per Definition \ref{def:weak:dist:min}.
	
	If program  ${\calDF}: I \rightarrow O$ is minimal, for $I' \subseteq I$, according to Definition \ref{def:mono-mini2}, then
	${\calDF}$ is also strong distributed minimal for $I'$ as per Definition \ref{def:dist:min}, and is thus also distributed minimal for $I'$.
\end{theorem}

\subsection{Monitoring mechanisms to detect strong distributed (non) minimality}
\label{sec:mon:dist}
Similar to the monolithic case, we are interested in
checking whether the inputs provided to an (untrusted) program $\calDF: I_1\times \cdots \times I_n \rightarrow O$ are minimized in the best possible way. We consider monitoring input-output behavior of program $\calDP$ where $\calDP$ is repeated execution of program $\calDF$.

By monitoring  ${\calDP}$ (input-output behavior of several executions of program ${\calDF}$), we are interested in checking whether an  execution of ${\calDP}$ satisfies strong distributed (non) minimality property.

\begin{figure}[t]
	\centering
	{\includegraphics[scale=0.8]{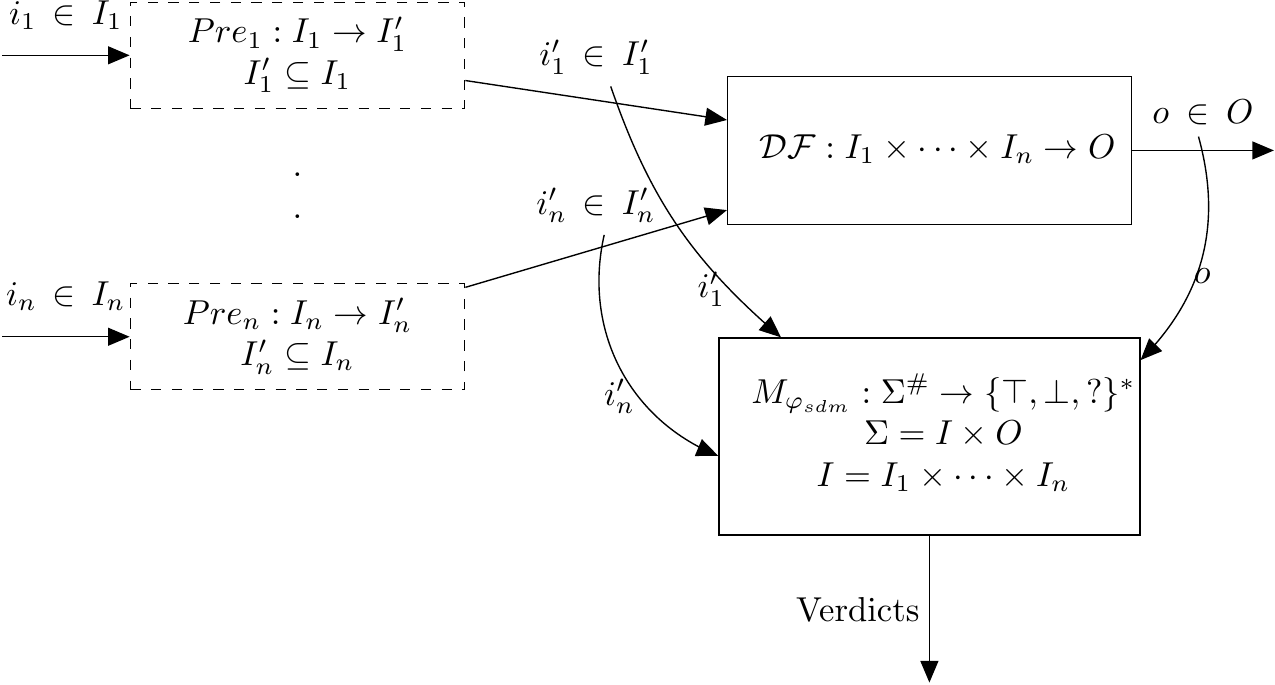}}
	\caption{Monitor $M_{dm}$ in the distributed case.}
	\label{fig:monitor:dist}
\end{figure}

The framework of the proposed monitoring approach in the distributed case, where the program has multiple input sources is depicted in Figure \ref{fig:monitor:dist}.
We consider that there are multiple input sources $[1,n]$, and the set of all possible values for each input source $id \in [1,n]$ is denoted using $I_{id}$.
Program ${\calDP}$ requires an input from all its input sources to produce an output.
An input event $i$ that the user (or $testInputGenerator$) provides belongs to the set $I$, ($i\in I = I_1\times\cdots\times I_n$).
We assume that inputs from the user are first pre-processed by a data pre-processor ($Pre$), and the pre-processed input that belongs to the set $I'$ (where $I' \subseteq I$) is fed as input to the \emph{untrusted}  program ${\calDF}$.
For each execution of ${\calDF}$, the monitor observes both the pre-processed input and the output of ${\calDF}$.

We consider that the monitor cannot observe and is not aware of the actual inputs $i_1,\cdots, i_n$ that the user provides.
Moreover, pre-processors may or may not exist.
We also assume that the monitor is unaware whether pre-processors exist or not, and that it does not know about their behavior.
 What the monitor observers at runtime is the pre-processed input $i'$ that belongs to $I'$ which also belong to $I$.

After each execution of program $\calDF: I_1\times\cdots\times I_n \rightarrow O$ (i.e., in every iteration of  program {$\calDP$}), the monitor observes the (pre-processed) input and the output of program $\calDF$ in that particular iteration (step) of {$\calDP$} .

For any word $\sigma\in \ws$ (current observation of execution of {$\calDP$}) of length greater than 1,  where $\Sigma = I \times O$ with $I_1\times\cdots\times I_n$, $M_{\varphi_{sdm}}$ as per Definition \ref{def:rv:mon} is a monitor for property $\varphi_{sdm}$.
The monitor returns $\true$ ($\top$) when $\sigma$ followed by any extension of it satisfies the distributed minimality property $\varphi_{sdm}$.
The monitor returns $false$ ($\bot$) when the current observation of execution of {$\calDP$} followed by any extension of it violates $\varphi_{sdm}$ (resp. satisfies $\overline{\varphi_{sdm}}$).
It returns $?$ (unknown) for the current observation if the other two cases do not hold.

Similar to the monolithic case,  checking whether the distributed minimality property is satisfied (resp. violated) for every extension of the current observation, needs to be reduced to checking whether the current observation satisfies (resp. violates) some constraints.
We now introduce the function $\satMinD$ that is defined based on definitions of properties $\varphi_{sdm}$ (resp. $\overline{\varphi_{sdm}}$).
$\satMinD$ is used to check whether the current observation $\sigma$ satisfies property $\varphi_{sdm}$ (resp. $\overline{\varphi_{sdm}}$) .

The function $\satMinD: \ws \rightarrow \bbb$, takes an input-output word $\sigma \in \ws$ and it returns a Boolean as output.
It is defined as follows:
\[
\begin{array}{lll}
\satMinD(\sigma)& =
\begin{cases}
\true & \mbox{if}\

\begin{array}{ll}
\forall i \in [ 1, |\sigma| ], \forall j\neq i \in [1, |\sigma| ], \\
\quad \text{ let } \Pi_1(\sigma_i)= (i_1, \cdots, i_n) \wedge  \Pi_1(\sigma_j) =(i'_1, \cdots, i'_n).\\
\quad \quad (\exists x \in [1,n]: i_{x} \neq i'_{x} \wedge \forall y\in [1,n]: y\neq x \implies i_{y}= i'_{y}) \\
\quad \quad\quad \quad\quad \quad \implies  \Pi_2(\sigma_i) \neq \Pi_2(\sigma_j).
\end{array}	\\
\false & Otherwise\\
\end{cases}
\end{array}
\]
$\satMinD$ checks whether a given word $\sigma$ belongs to property $\varphi_{sdm}$. For any given word $\sigma$, $\satMinD(\sigma)$ is $\true$ if $\sigma \in \varphi_{dm}$, and is $\false$ otherwise, i.e., if it returns $\false$, then $\sigma \in \overline{\varphi_{sdm}}$.


Similar to the monolithic case, the condition for the second case of the monitor for property $\varphi_{sdm}$ can be simplified.
\begin{proposition}
	\label{prop:false:dist}
	Given any word $\sigma \in \ws$, where $|\sigma| >1$, we have that
	$\text{ if } \sigma \in \overline\varphi_{sdm} \text{ then } (\forall \sigma': \sigma \pref \sigma', \sigma' \in \overline{\varphi_{sdm}}).$
\end{proposition}
\begin{remark}
	\label{prop1:true:dist}
Similar to the monolithic case, regarding the condition of the $\top$ case (satisfaction of $\varphi_{sdm}$), checking whether the current observed word $\sigma$ belongs to $\varphi_{sdm}$ (i.e.,whether $\satMinD(\sigma)$ is $\true$) is not sufficient, as it does not ensure that every extension of $\sigma$ will also belong to $\varphi_{sdm}$.
Thus, testing condition of the first case is not possible in general.
\end{remark}
However, by providing additional knowledge to the monitor about all the input domains, and when they are bounded, checking the condition of the first case reduces to  testing whether the current observation satisfies $\varphi_{sdm}$, and if it also covers all possible inputs.
\begin{proposition}
	\label{prop:true:dist}
	Given any word $\sigma \in \ws$, where $\Sigma = I \times O$, with $I = I_1\times\cdots I_n$, and when $|\sigma| >1$, we have that
	$\text{ if }(\inpExh(I, \sigma) \wedge \satMinD(\sigma) ) \text{ then } (\forall \sigma': \sigma \pref \sigma', \sigma' \in {\varphi_{sdm}}).$
%
\end{proposition}
%
Using Propositions \ref{prop:false:dist} and \ref{prop:true:dist}, the conditions of the first two cases in $M_{\varphi_{sdm}}$ can be simplified. We present the simplified definition below, where the conditions of the first two cases are reduced to checking whether the observed input word satisfies some constraints.

Consider the property $\varphi_{sdm} \subseteq \ws$, where $\Sigma= I \times O$, and $I = I_1\times\cdots \times I_n$.
Let $\sigma \in \ws$ denote a finite input-output word over the alphabet $\Sigma = I \times O$ (current observation of an execution of $\calDP$ which belongs to $\calL(\calDP)$).
The monitor for strong distributed minimality is denoted as $M_{\varphi_{sdm}}$.
For $\sigma = \epsilon$ and any word $\sigma$ of length 1, $M_{\varphi_{sdm}}(\sigma) = ?$.
%
  $M_{\varphi_{sdm}}$ is defined as follows:

\begin{definition}[Monitor for strong distributed minimality]
	\label{def:monitor:dist}
	A monitor for  property $\varphi_{sdm}$  is a function $M_{\varphi_{sdm}}: \ws \rightarrow \D$, where $D = \{\top, \bot, ?\}$ is defined as follows:
%
	\[
	\begin{array}{lll}
	M_{\varphi_{sdm}}(\sigma)& =
	\begin{cases}
	\top & \mbox{if}\ |\sigma>1| \wedge \inpExh(I, \sigma) \wedge \satMinD(\sigma)\\
	\bot & \mbox{if}\ |\sigma>1| \wedge \neg\satMinD(\sigma)\\
	? & Otherwise
	\end{cases}
	\end{array}
	\]
\end{definition}

\begin{table}[t]
	\centering
	\begin{tabular}{|c|c|}
		\hline
		\multicolumn{1}{|c|}{$\sigma$} & \multicolumn{1}{c|}{$M(\sigma)$} \\
		\hline
		{$((5000, 45),\true)$} & $?$  \\
		\hline
		$((5000, 45),\true) \cdot ((11000, 51),\false)$ & $?$ \\
		\hline
		$((5000, 45),\true) \cdot ((11000, 51),\false) \cdot ((4000, 21),\true)$ & $?$ \\
		\hline
		$((5000, 45),\true) \cdot \bm{((11000, 51),\false)} \cdot ((4000, 21),\true)\cdot \bm{((11000, 55), \false)}$  & $\bot$ \\
		\hline
	\end{tabular}
	\caption{Example illustrating behavior of monitor $M_{dm}$.}
	\label{tableExample2}
\end{table}

\begin{example}
	Let us consider the example program $\calP$  illustrated in Figure~\ref{fig:prog:dist}.
	In Table \ref{tableExample2}, we present some example observations of an execution of program $\calDP$ being monitored denoted as $\sigma$, and the verdict provided by the monitor for $\sigma$.
\end{example}

\begin{proposition}
	\label{prop:monitor:dist}
	$M_{\varphi_{sdm}}$ in Definition \ref{def:monitor:dist} is a monitor for  property $\varphi_{sdm}$ as per Definition \ref{def:rv:mon}.  
\end{proposition}

\section{Pre-deployment testing and minimiser synthesis via monitoring}
\label{sec:preDeployment}
The discussion and results of this section applies to both the monolithic and distributed cases. 
To simplify the presentation, we illustrate and discuss the results considering monitoring of monolithic minimality.

Let us consider the definition of monitor $M_{\varphi_m}$ (Definition \ref{def:monitor:mono2}). 
In order to provide conclusive verdict $\top$ from an observed input-output word $\sigma \in \ws$,
in addition to testing $\satMin(\sigma)$, the monitor has to be provided with information about the set of all possible inputs $I$, and we need to check whether every possible input appear in $\sigma$ at least once (i.e., test whether $\inpExh(I, \sigma)$ holds).

In runtime monitoring, the word $\sigma \in \ws$ (observation of current execution of $\calP$) that is fed to a monitor is of finite bounded length ($\sigma$ and its length are both known).  
Thus, for any given $\sigma\in \ws$ and any set of inputs $I$, testing $\inpExh(I, \sigma)$ is straightforward, as illustrated in
Algorithm \ref{algo:inpExt}.
%
\begin{algorithm}[ht]
	\caption{$ \inpExh(I, \sigma) $}
	\label{algo:inpExt}
	{
		\begin{algorithmic}[1]
			\STATE $I' \gets \{ \} $
			\FOR{$ i \in [ |1, |\sigma| ] $}
			\STATE $inp \gets \Pi_1(\sigma_i)$
			\STATE $I' \gets I' \cup \{ inp \}$
			\IF{$|I'| = |I|$}
			\RETURN $\true$
			\ENDIF		
			\ENDFOR
			\RETURN $\false$
		\end{algorithmic}
	}
\end{algorithm}

Algorithm \ref{algo:inpExt} ($\inpExh$) requires the set of possible inputs $I$, and an input-output word $\sigma \in \ws$ (where $\Sigma = I \times O$) as input parameters. $I'$ which is initially empty is used to keep track of the set of inputs seen in $\sigma$.
While processing the sequence $\sigma$ event by event to build $I'$, if
$|I'| = |I|$, then the algorithm returns $\true$ and terminates.
In the worst-case, the for-loop runs for $|\sigma|$ times.  
After processing all the events in $\sigma$, if $|I'| \neq |I|$ then the algorithm returns $\false$.

Note that before entering the for-loop in Algorithm  \ref{algo:inpExt} it is checked whether $|\sigma| < |I|$. If so, we can immediately return $\false$.
\begin{proposition}[$|\sigma| < |I|$]
	\label{prop:inpLength}
	When the length of the input-output word $\sigma \in \ws$
	is less that the cardinality of the set of inputs, then $\inpExh(I, \sigma)$ is $\false$:
	\[\forall \sigma\in \ws: |\sigma| < |I| \implies \inpExh(I, \sigma) = \false\]
\end{proposition}
%
\begin{remark}[Conclusive verdict $\top$ during runtime monitoring]
In general, when performing online monitoring of program $\calP$ (where $\sigma \in \ws$ is the current observation of execution of $\calP$), and providing knowledge of the set of all possible inputs of $\calP$ to the monitor, it is highly unlikely that $\sigma$ covers all the inputs in $I$. 
Thus, as we can imagine, during runtime monitoring
$ \inpExh(I, \sigma)$ most likely will return $\false$, and thus the condition of the first case (that provides conclusive verdict $\top$) in Definition \ref{def:monitor:mono2} most likely does not hold, and  thus we notice only verdicts $?$ or $\bot$ in practice.
\end{remark}

However, monitor $M_{\varphi_m}$ can be used for testing for minimality prior to deployment. In this case, the input observation fed to the monitor can be generated in such a way that it covers all the inputs in $I$ (when $I$ is finite and bounded).
\subsection{Testing in a controlled environment via monitoring}
%
When the set of inputs $I$ is bounded, and when testing $\calF$ in a controlled environment (i.e., when we have control over the inputs that are fed to $\calF$), it is indeed possible to obtain a conclusive verdict (either $\top$ or $\bot$), upon observing a sequence $\sigma$ of length $|I|$.
We discuss this further via the monolithic case, which straightforwardly extends to the distributed case.

%
\begin{algorithm}[ht]
	\caption{$\mathsf{DataMinTester}$}
	\label{algo:minTester}
	{
		\begin{algorithmic}[1]
			\STATE $I' \gets I$, $\sigma \gets \epsilon$, $v \gets ?$
			\WHILE {($|I'| > 0 \wedge v==?$)}
			\STATE $i \gets pickInp(I')$
			\STATE $o \gets \calF(i)$
			\STATE $\sigma \gets \sigma \cdot(i,o)$
			\STATE $v \gets M_{\varphi_m}(\sigma)$
			\STATE $I' \gets I' \setminus \{i\}$
			\ENDWHILE
		\end{algorithmic}
	}
\end{algorithm}
%

Algorithm \ref{algo:minTester} (DataMinTester) is for testing $\calF$ via monitoring.
In Algorithm \ref{algo:minTester}, $I'$ contains inputs that are not yet fed to the program $\calF$.
Initially, $I'$ is assigned with the set of inputs $I$.
In every iteration of the while loop, an input $i$ from the set $I'$ is picked non-deterministically, which is fed to $\calF$, and $\calF(i)$ is assigned to $o$. The input-output event $(i,o)$ is then fed to the monitor.
Before proceeding to the next iteration, input $i$ which is already considered in the current iteration is removed from the set $I'$.

Algorithm \ref{algo:minTester} (DataMinTester) can be considered as program $\calP$ where  program $\calF$ is executed repeatedly.
However, here, in every iteration we invoke $\calF$ with a new input from the set $I$ (i.e., input that has not been considered in the previous iterations).
The while loop thus terminates after $|I|$ iterations.

After $|I|$ iterations of the algorithm, the input-output word $\sigma$ that the monitor receives will be of length $|I|$, and $\inpExh(I, \sigma)$ will evaluate to $\true$.
The monitor $M_{\varphi_m}$ certainly returns a conclusive verdict upon receiving an input-output word of length $|I|$.
%
\begin{proposition}
Let $\sigma \in \ws$ be an execution of the $\mathsf{DataMinTester}$ (Algorithm \ref{algo:minTester}), which is a sequence of input-output word fed to the monitor.
The length of $\sigma$ will be at most $|I|$, and
 the monitor will certainly return a conclusive verdict $\top$ or $\bot$ for  $\sigma$ of length $|I|$.
\end{proposition}
Regarding conclusive verdict $\top$ (i.e, about satisfaction of minimality), the monitor cannot provide this verdict before observing a word of length $|I|$.
Regarding conclusive verdict $\bot$ (i.e, about violation of minimality), the monitor may be able to provide this verdict before observing word of length $|I|$.
In this case, the execution of the tester program can stop earlier soon after the monitor observes the sequence that satisfies non-minimality property.
%

\subsection{Minimiser synthesis}
\label{sec:minimiser:synthesis}
In this section, by considering the monolithic case, we briefly present about possibility of synthesizing a pre-processor for $\calF$ such that the composition of $\calF$ with the synthesized pre-processor satisfies the data minimality principle.

An input pre-processor that does not change the behavior of $\calF$, and makes $\calF$ minimal when composed with it is called as a minimiser, defined as follows: 
 \begin{definition}[Minimizer]
 	\label{def:minimiser}
 	A pre-processor $Pre: I \rightarrow I$  is a monolithic \emph{minimiser} for $\calF$ iff $Pre$ is a pre-processor for $\calF$ and $\calF$ is monolithic-minimal for $\range(Pre)$.
 \end{definition}
%
 \begin{figure}[t]
 	\centering
 	{\includegraphics[scale=0.7]{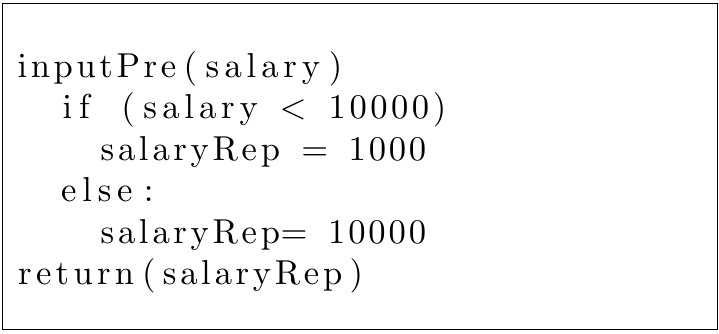}}
 	\caption{Example of an input pre-processor (which is also a minimiser) for program $\calF$ illustrated in Figure\ref{fig:func:mono}.}
 	\label{fig:eg:minimiser}
 \end{figure}
\begin{example}
The input pre-processor presented in Figure \ref{fig:eg:pre} is not a minimiser for program $\calF$  illustrated in Figure\ref{fig:func:mono}, since we can define a pre-processor with two cases. The input pre-processor presented in Figure \ref{fig:eg:minimiser} is also a minimiser for $\calF$ in Figure\ref{fig:func:mono}.
\end{example}

We show that when the input domain $I$ is bounded and known, by monitoring input-output behavior of $\calF$, in addition to checking whether $\calF$ is minimal (resp. non-minimal), it is also possible to synthesize a minimiser.
\paragraph{Algorithm for obtaining a minimiser.}
Algorithm \ref{algo:minTester} can be adapted for building a partitioning of the input domain $I$, where for every partition, program $\calF$ produces the same output for any input belonging to that partition.
The condition of the while-loop should now be $|I'| >0$ since we  need to continue the execution (irrespective of whether the monitor provides a conclusive verdict), to cover all the inputs to build a partitioning of the input domain.
In each iteration of the while-loop:
\begin{itemize}
\item Let $i$ be the input picked from the set of un-covered inputs, and let $o$ be the output produced by $\calF$. 
\item If an input partition corresponding to $o$ already exists, then $i$ is added to that input partition. Otherwise, a new partition corresponding to output $o$ is created before proceeding to the next iteration.  
\item The while-loop terminates after $|I|$ iterations, and we have a partitioning of the input domain.
\end{itemize}
 For each input partition, an element is chosen (non-deterministically) as input representative for that partition.
$I'\subseteq I$ is the set of input representatives, and the algorithm returns a mapping from $I$ to $I'$, where for each input partition, every element belonging to that partition is mapped to its corresponding input representative.
\begin{proposition}
Consider any program $\calF: I \rightarrow O$,  
When $I$ is known and $|I|$ is bounded, the algorithm for obtaining a minimiser discussed above terminates and it returns a minimizer for program $\calF$.
\end{proposition}

\section{Implementation}
\label{sec:impl}

The runtime monitoring mechanisms for checking minimality (resp. non-minimality) for both the monolithic and distributed cases have been implemented in Python.
The main goal of this prototype implementation is to validate the feasibility and practicality of the proposed approaches (i.e., monitoring for (non) minimality at runtime, pre-deployment testing and synthesis of a minimiser).

\paragraph{Implementation of monitors.}
Regarding the implementation of monitors (e.g., the implementation of $M_{\varphi_m}$), implementation of functions $\satMin$ (resp. $\satNonMin$) that checks whether a given trace (current observation of an execution of $\calP$)  is minimal (resp. non-minimal) is straightforward from their definitions.
For the first case ($\top$) in the definition of $M_{\varphi_m}$, we also additionally need to check whether the current observation $\sigma$ covers all the inputs (i.e., whether $inpExh(I, \sigma)$ holds, where $I$ is the set of all possible inputs).

\paragraph{Monitoring for (non) minimality at runtime.}
For testing the usage of monitors at runtime for detecting (non) minimality, 
we wrapped the program (to be monitored) with a user simulator (test-input generator).
The user simulator is an (infinite) while-loop, where in each iteration, 
the program (being monitored) is invoked with some input $i$ chosen non-deterministically from the set of allowed inputs $I$, and the monitor is fed with the input $i$ and the output $o$ that the program returns.
The loop terminates when the monitor returns a conclusive verdict ($\top$ or $\bot$).
When the monitor returns conclusive verdict $\bot$, it also returns an evidence that shows violation of data minimality. 

For example, when the program for computing benefits illustrated in Figure \ref{fig:func:mono} composed with the input pre-processor illustrated in Figure \ref{fig:eg:pre} is considered as the program to be monitored, as expected the approach terminated and returned conclusive verdict $\bot$ with an evidence.  

When the program for computing benefits in Figure \ref{fig:func:mono} composed with the minimiser in Figure \ref{fig:eg:minimiser} is considered as the program to be monitored, we can notice that the user simulator does not terminate. As expected, coverage of all inputs is highly impossible in this approach and the monitor always returns verdict unknown (?).
 
\paragraph{Pre-deployment testing.} 
We discussed about using the monitor to test the program in a controlled environment (Algorithm \ref{algo:minTester}), and check whether it satisfies the data minimality principle.
Algorithm \ref{algo:minTester} also has been implemented and tested.

When the program for computing benefits in Figure \ref{fig:func:mono} composed with the minimiser in Figure \ref{fig:eg:minimiser} is considered as the program to be tested, as expected, the approach returned conclusive verdict $\top$ (i.e., the composition of the program in Figure \ref{fig:func:mono} with the minimiser in Figure \ref{fig:eg:minimiser} satisfies data minimality).  

\paragraph{Minimiser synthesis.}
The algorithm for synthesizing a minimiser discussed briefly in Section \ref{sec:minimiser:synthesis} has been also implemented for the monolithic case. 
When the program for computing benefits in Figure \ref{fig:func:mono} is considered with $I = \{1, \cdots, 30000\}$, the minimiser synthesizer returned a minimiser where partitioning of $I$ consists of two partitions. 
The set $\{1, \cdots, 9999\}$ is one partition and all the elements in this set are mapped to a representative chosen from it (e.g., $9999$), and the set $\{10000, \cdots, 30000\}$ is the other partition, and all the elements from this set are mapped to an input representative chosen from it (e.g., $30000$).  

\begin{figure}[t!]
	\begin{center}
\begin{lstlisting}[language=Python]
##############################################
## input numFlights: integer between 0 to 100#
##############################################
def computeStatusLevel(numFlights):
    status = 0
    i = 0
    if numFlights<10:
        status = 0
    elif numFlights<20:
        status = numFlights -10
    elif numFlights<30:
        while i<= numFlights-20:
            status = status+numFlights
            i=i+1
        if status>150:
            status = 150
    else:
        status = 500
    return status
##############################################

\end{lstlisting}
	\end{center}
	\caption{Program $\calF$ to compute a loyalty status.}
	\label{loyaltyapp}
\end{figure}
\begin{example}
We consider the following example from \cite{ASS17dm}. 	
An airport facility must provide services to customers depending on their status.
The status level of a customer is determined depending on the number of flights taken by the customer in the previous year with its favorite company, \emph{PrivaFly}.
This number of flights information is disclosed by the airline company to the airport.

However, \emph{PrivaFly} wants to adopt the best practices in personal data protection, and requires only the needed data to be disclosed.
The airport services have their own policy to compute the status level, program $\calF$ shown in Figure~\ref{loyaltyapp}.
The program $\texttt{compStatusLevel}$ takes information about the number fo $\texttt{flights}$, and it returns a $\texttt{status}$ level.
If the number of flights is lower than $9$, the status is $0$. From $10$ to $19$ flights, the status level is $\texttt{numFlights}$-10.
If the number of flights ranges from $20$ to $29$, the computation status involves a loop.
Finally, over $30$ flights, the status level is capped to $500$.

Intuitively, there is no need to give a precise value for a number of flights between $0$ and $9$ and over $30$.
On the other hand, the exact number should be disclosed between $10$ and $19$.

The minimiser synthesis approach partitions the set of possible inputs (integer between 1 to 100) into 17 partitions, i.e., the set of possible outputs of the minimiser consists of 17 elements.
Elements in $[0,10]$ are grouped into a partition, and every element in this partition is mapped to a representative chosen from this partition. For elements between 11 to 24, there will be 14 partitions each partition consisting of one element. 
Elements in $[25,29]$ are grouped into one partition (due to sealing of the status to 150 for flights between 20 to 29, the program returns the same output for input value between 25 and 29).
All the remaining elements in $[30,100]$  are grouped  into a partition. 
The approach returns a minimiser in less than $0.02$ seconds.
\end{example}
\begin{remark}[Minimiser synthesis approach in \cite{ASS17dm}]
The example illustrated in Figure \ref{loyaltyapp}, is one of the examples provided with the implementation described in \cite{ASS17dm}.
Note that the minimiser synthesis approach in \cite{ASS17dm} requires the source code of the program. 
In our monitoring approach, the program can be a black-box, in the example considered above, we only need to know that program $\texttt{compStatusLevel}$ requires an integer between 1 to 100 (number of flights taken), and it returns information about the status. 
The approach in \cite{ASS17dm} is also very complex, involving symbolic execution of the program and the use of SAT solver to obtain a partitioning of the input space.  
Moreover, when the program contains loops (e.g., $\texttt{compStatusLevel}$ program in Figure \ref{loyaltyapp}) the approach in \cite{ASS17dm} cannot synthesize a minimiser in general. By adding loop-invariants (that helps the symbolic executor to handle loops), the approach may generate a pre-processor, and whether the generated pre-processor is a minimiser or not depends on the added loop-invariants.
\end{remark}

\section{Conclusion and Future Work}
\label{sec:conclusion}

Data minimisation is a privacy enhancing principle, stating that personal data collected should be no more than necessary for the specific purpose consented by the user.
The data minimisation process aims to minimise the input data such that only data that is necessary is given to the program.

In this paper, we consider the problem of runtime monitoring of deterministic programs to detect (non) minimality.
We propose monitoring mechanisms where a monitor observes the inputs and the outputs of a program, to detect violation of data minimisation policies.
We formally define runtime monitors to check (non-)minimality for both the monolithic and the distributed case.
We show that checking for satisfaction of minimality (i.e., giving a conclusive verdict $\top$ for satisfaction of minimality) via monitoring is not possible in general for any of the cases, and that non-minimality for the monolithic and a strong version of the distributed cases can be checked in general, but not for the normal distributed case.
We prove that under certain conditions we can monitor and check both minimality and its negation for all cases.
We describe a procedure that gives a definite answer on whether the program is minimal or not by using runtime monitoring in a controlled (pre-deployment) test environment, and also obtain a minimiser for the program under test.
The proposed results for both online and offline monitoring have been implemented as a proof-of-concept.

In the near future, we plan to generalise the monitoring results discussed in this paper for other security policies for deterministic programs.
We also intend to study and formalise the concept of data minimisation for non-deterministic systems, and explore on the monitoring and minimiser synthesis problem for such systems.

\bibliographystyle{splncs}
\bibliography{biblio}

\begin{thebibliography}{10}
\providecommand{\url}[1]{\texttt{#1}}
\providecommand{\urlprefix}{URL }

\bibitem{ASS17dm}
Antignac, T., Sands, D., Schneider, G.: {Data Minimisation: A Language-Based
  Approach}. In: IFIP Information Security \& Privacy Conference (IFIP SEC'17).
  IFIP Advances in Information and Communication Technology (AICT), vol. 502,
  pp. 442--456. Springer Science and Business Media (2017)

\bibitem{Bauer:2011:RVL}
Bauer, A., Leucker, M., Schallhart, C.: Runtime verification for {LTL} and
  {TLTL}. ACM Trans. Softw. Eng. Methodol.  20(4),  14:1--14:64 (Sep 2011),
  \url{http://doi.acm.org/10.1145/2000799.2000800}

\bibitem{Blech2012}
Blech, J.O., Falcone, Y., Becker, K.: Towards certified runtime verification.
  In: Aoki, T., Taguchi, K. (eds.) Formal Methods and Software Engineering:
  14th International Conference on Formal Engineering Methods, ICFEM 2012,
  Kyoto, Japan, November 12-16, 2012. Proceedings. pp. 494--509. Springer
  Berlin Heidelberg, Berlin, Heidelberg (2012)

\bibitem{Bonakdarpour2016}
Bonakdarpour, B., Finkbeiner, B.: Runtime Verification for HyperLTL, pp.
  41--45. Springer International Publishing, Cham (2016)

\bibitem{hyp1}
Clarkson, M.R., Schneider, F.B.: Hyperproperties. J. Comput. Secur.  18(6),
  1157--1210 (Sep 2010),
  \url{http://dl.acm.org/citation.cfm?id=1891823.1891830}

\bibitem{cohen1977}
Cohen, E.: Information transmission in computational systems. SIGOPS Oper.
  Syst. Rev.  11(5),  133--139 (Nov 1977),
  \url{http://doi.acm.org/10.1145/1067625.806556}

\bibitem{ColomboPS08}
Colombo, C., Pace, G.J., Schneider, G.: Dynamic event-based runtime monitoring
  of real-time and contextual properties. In: Cofer, D.D., Fantechi, A. (eds.)
  Formal Methods for Industrial Critical Systems, 13th International Workshop,
  {FMICS} 2008, L'Aquila, Italy, September 15-16, 2008, Revised Selected
  Papers. Lecture Notes in Computer Science, vol. 5596, pp. 135--149. Springer
  (2008), \url{https://doi.org/10.1007/978-3-642-03240-0}

\bibitem{DIEKERT201429}
Diekert, V., Leucker, M.: Topology, monitorable properties and runtime
  verification. Theoretical Computer Science  537(Supplement C),  29 -- 41
  (2014), iCTAC 2011

\bibitem{gdpr2016}
{European Parliament and Council}: {Regulation (EU) 2016/679 of the European
  Parliament and of the Council of 27 April 2016 on the protection of natural
  persons with regard to the processing of personal data and on the free
  movement of such data, and repealing Directive 95/46/EC (General Data
  Protection Regulation)} (apr 2016)

\bibitem{FalconeFM09}
Falcone, Y., Fernandez, J., Mounier, L.: Runtime verification of
  safety-progress properties. In: Bensalem, S., Peled, D.A. (eds.) Runtime
  Verification, 9th International Workshop, {RV} 2009, Grenoble, France, June
  26-28, 2009. Selected Papers. Lecture Notes in Computer Science, vol. 5779,
  pp. 40--59. Springer (2009),
  \url{http://dx.doi.org/10.1007/978-3-642-04694-0}

\bibitem{Falcone2009}
Falcone, Y., Fernandez, J.C., Mounier, L.: Runtime Verification of
  Safety-Progress Properties, pp. 40--59. Springer Berlin Heidelberg (2009)

\bibitem{Finkbeiner2015}
Finkbeiner, B., Rabe, M.N., S{\'a}nchez, C.: Algorithms for Model Checking
  HyperLTL and HyperCTL, pp. 30--48. Springer International Publishing, Cham
  (2015)

\bibitem{Havelund2008}
Havelund, K., Goldberg, A.: Verify your runs. In: Verified Software: Theories,
  Tools, Experiments: First IFIP TC 2/WG 2.3 Conference, VSTTE 2005, Zurich,
  Switzerland, October 10-13, 2005, Revised Selected Papers and Discussions.
  pp. 374--383. Springer Berlin Heidelberg, Berlin, Heidelberg (2008)

\bibitem{DBLP:conf/setss/Leucker16}
Leucker, M.: Runtime verification for linear-time temporal logic. In: Bowen,
  J.P., Liu, Z., Zhang, Z. (eds.) Engineering Trustworthy Software Systems -
  Second International School, {SETSS} 2016, Chongqing, China, March 28 - April
  2, 2016, Tutorial Lectures. Lecture Notes in Computer Science, vol. 10215,
  pp. 151--194 (2016), \url{https://doi.org/10.1007/978-3-319-56841-6}

\bibitem{LeuckerS08jlap}
Leucker, M., Schallhart, C.: A brief account of runtime verification. Journal
  of Logic and Algebraic Programming  78(5),  293--303 (may/june 2009),
  \url{http://dx.doi.org/10.1016/j.jlap.2008.08.004}

\bibitem{oecd2013}
{Organisation for Economic Co-operation and Development}: {The OECD Privacy
  Framework}. Guidelines, Organisation for Economic Co-operation and
  Development (2013), chapter 1. Recommendation of the Council concerning
  Guidelines governing the Protection of Privacy and Transborder Flows of
  Personal Data (2013)

\bibitem{Pnueli2006}
Pnueli, A., Zaks, A.: PSL Model Checking and Run-Time Verification Via Testers,
  pp. 573--586. Springer Berlin Heidelberg, Berlin, Heidelberg (2006)

\bibitem{smith2009}
Smith, G.: On the foundations of quantitative information flow. In: de~Alfaro,
  L. (ed.) Foundations of Software Science and Computational Structures,
  Lecture Notes in Computer Science, vol. 5504, pp. 288--302. Springer Berlin
  Heidelberg (2009)

\bibitem{fipp1973}
{US Secretary's Advisory Committee on Automated Personal Data Systems}:
  {Records, Computers and the Rights of Citizens}. Report DHEW NO. (OS)73-94,
  US Secretary's Advisory Committee on Automated Personal Data Systems,
  Brussels, Belgium (July 1973), chapter IV: Recommended Safeguards for
  Administrative Personal Data Systems

\end{thebibliography}
\end{document}